\documentclass[%
 reprint,
 amsmath,amssymb,
 aps,
]{revtex4-1}

\usepackage{color}

\usepackage{hyperref}
\usepackage{graphicx}
\usepackage{dcolumn}
\usepackage{bm}

\newcommand{\beginsupplement}{%
        \setcounter{table}{0}
        \renewcommand{\thetable}{S\arabic{table}}%
        \setcounter{figure}{0}
        \renewcommand{\thefigure}{S\arabic{figure}}%
}

\begin{document}

\preprint{APS/123-QED}

\title{Optimally controlling the human connectome: the role of network topology}

\author{Richard F. Betzel$^1$}
\author{Shi Gu$^1$}
\author{John D. Medaglia$^{1,2}$}
\author{Fabio Pasqualetti$^3$}
\author{Danielle S. Bassett$^{1,4}$}
 \email{dsb @ seas.upenn.edu}
\affiliation{
 $^1$Department of Bioengineering, University of Pennsylvania, Philadelphia, PA, 19104
}
\affiliation{
 $^2$Department of Psychology, University of Pennsylvania, Philadelphia, PA, 19104
}
\affiliation{
 $^3$Department of Mechanical Engineering, University of California, Riverside, Riverside, CA, 92521
}
\affiliation{
$^4$Department of Electrical and Systems Engineering, University of Pennsylvania, Philadelphia, PA, 19104
}

\date{\today}

\begin{abstract}
To meet ongoing cognitive demands, the human brain must seamlessly transition from one brain state to another, in the process drawing on different cognitive systems. How does the brain's network of anatomical connections help facilitate such transitions? Which features of this network contribute to making one transition easy and another transition difficult? Here, we address these questions using network control theory. We calculate the optimal input signals to drive the brain to and from states dominated by different cognitive systems. The input signals allow us to assess the contributions made by different brain regions. We show that such contributions, which we measure as energy, are correlated with regions' weighted degrees. We also show that the network communicability, a measure of direct and indirect connectedness between brain regions, predicts the extent to which brain regions compensate when input to another region is suppressed. Finally, we identify optimal states in which the brain should start (and finish) in order to minimize transition energy. We show that the optimal target states display high activity in hub regions, implicating the brain's rich club. Furthermore, when rich club organization is destroyed, the energy cost associated with state transitions increases significantly, demonstrating that it is the richness of brain regions that makes them ideal targets.
\end{abstract}

\maketitle


\section{Introduction}

One of the goals of modern biology is to understand how a system's form influences its functionality. Human brain networks manifest structure-function relationships, with converging evidence suggesting that the brain's network of white-matter fiber pathways (\emph{structural connectivity}; SC) constrains the intrinsic functional interactions among brain regions at rest (\emph{functional connectivity}; FC), thereby shaping the emergence of coherent spatiotemporal patterns of neural activity \cite{hagmann2008mapping, honey2009predicting, hermundstad2013structural,atasoy2016human,becker2015accurately}. The effect of these constraints is that over long periods of time (hours, days) resting FC largely recapitulates the underlying SC \cite{deco2011emerging} so that the strongest functional interactions are often mediated by direct anatomical projections \cite{goni2014resting}. Over shorter timescales, however, FC is more variable, decoupling from the underlying anatomy to engage specific cognitive systems both at rest \cite{allen2012tracking, karahanouglu2015transient, betzel2016dynamic} and in order to meet ongoing cognitive demands \cite{bassett2011dynamic, braun2015dynamic, mattar2015functional}.

How does the brain smoothly transition from the activation of one cognitive system to the activation of another?  What are the anatomical and topological substrates that facilitate such transitions? One approach for addressing these and similar questions is to model the human brain as a dynamical system, treating brain regions as dynamic elements with time-dependent internal states. As the system evolves, each brain region's state is updated according to its own history and the states of its connected neighbors. Such models vary in their complexity and neurophysiological basis, ranging from biophysically plausible descriptions of interacting neuronal populations \cite{wilson1972excitatory, deco2008dynamic, marreiros2010dynamic, freyer2011biophysical, muldoon2016stimulation} to abstract models based on oscillations, diffusion, and epidemic spreading \cite{breakspear2010generative, haimovici2013brain, abdelnour2014network, misic2015cooperative}.

In most applications, the question of how to control distributed brain dynamics is not explicitly considered. Here, control refers to the possibility of manipulating a dynamical system so that it evolves to follow a particular trajectory through its state space. We posit that the nature of brain state transitions can be meaningfully addressed with network control theory, which offers a mathematical framework for studying and, ultimately, controlling the evolution of dynamical systems on networks \cite{kalman1963mathematical, liu2011controllability}.

Most dynamical systems can be framed in a control perspective by introducing exogenous input to the system through a set of control sites (network nodes) in the form of time-varying signals. The effect of such inputs is to drive the system along a trajectory through its state space; different inputs, then, result in different trajectories \cite{liu2011controllability, sun2013controllability}. The effect of input on a system's trajectory depends upon (i) the system's dynamics, (ii) the composition of the control sites, and (iii) the configuration of the system's nodes and edges into a network (its topology)  \cite{liu2011controllability}. Understanding how control occurs in the brain and how these factors contribute to enacting control is of critical importance, with clear clinical and engineering implications. For example, the efficacy of implantable neuromodulatory devices for suppressing Parkinsonian symptoms \cite{rosin2011closed}, seizure abatement in epilepsy \cite{taylor2015optimal}, and other methods for manipulating brain activity, such as transcranial magnetic stimulation \cite{fox2012measuring}, depend on our ability to modify a network's function by introducing external electromagnetic signals.

In the current work we use network control theory to identify minimum energy input signals that cause the system to transition to and from specific brain states. The input signals are introduced through control sites -- brain regions -- and can be thought of in two complementary ways. One view is that control signals are issued directly from brain regions, themselves acting as local computational elements administering control over the network. Alternatively, control signals can be viewed as having extracranial provenance, originating from implanted electrodes or other neuromodulatry devices and thereby acting on specific brain regions. In either case, a region's local contribution can be modeled as an input of energy over time and interpreted as a measure of the amount of effort it puts forth during a control task \cite{yan2012controlling, sun2013controllability}. 

We seek to better understand the role of brain network topology in determining a region's energy -- what topological factors contribute to making transitions between different brain states more or less effortful? Previous investigations addressing similar questions have focused on how the underlying networks' statistical properties (e.g. the shape of degree distribution) and global metrics (e.g. modularity, clustering, etc.) influence the minimum number of control sites necessary to render the network controllable \cite{liu2011controllability, wang2012optimizing, cowan2012nodal, posfai2013effect, ruths2014control}. Such approaches, while illuminating, are limited. First, the focus on global network statistics makes it difficult to assess the contributions of individual nodes or edges. Second, the classification of networks as either ``controllable'' or ``uncontrollable'' overlooks finer gradations in the amount of energy required for control. Though some node-level metrics have been proposed \cite{pasqualetti2014controllability}, the precise roles of individual nodes and other topological features in facilitating control is not well understood. Previous applications of network control theory to brain networks investigated related questions, by studying \emph{all possible state transitions} and assuming an infinite time horizon \cite{gu2015controllability, muldoon2016stimulation}. Here, we focus on finite-time transitions between a limited set of accessible states, which we choose to correspond to previously-defined brain systems \cite{yeo2011organization}. 

\begin{figure*}[t]
\begin{center}
\centerline{\includegraphics[width=1\textwidth]{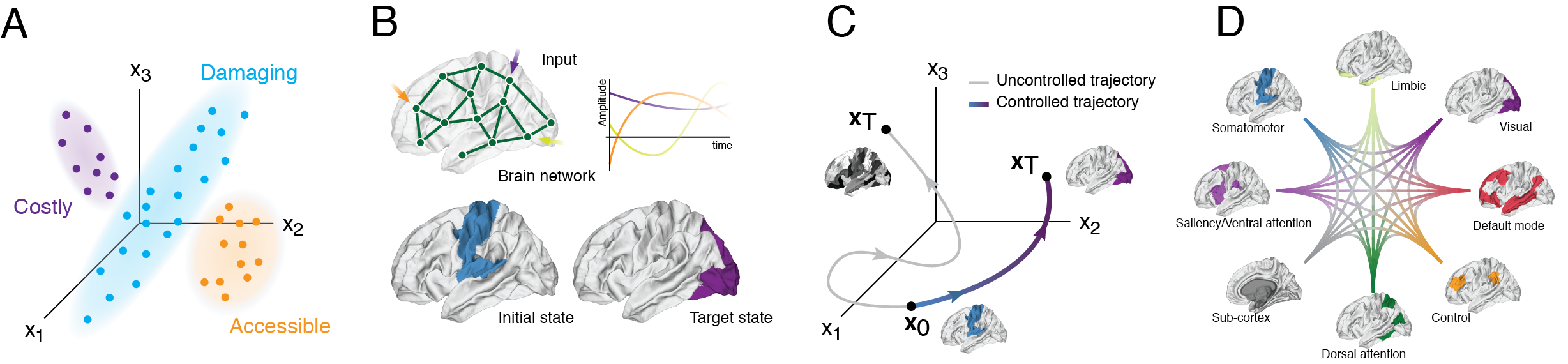}}
\caption{\textbf{Schematic illustrating structural controllability framework.} \emph{(A)} The space of all possible states includes regions that require prohibitively large energy to access as well as damaging configurations. Our analyses are restricted to the accessible region of state space populated by viable configurations. \emph{(B)} A set of time-varying inputs are injected into the system at different control points (nodes; brain regions). The aim is to drive the system from some particular initial state to a target state (e.g. from somatosensory to visual system). \emph{(C)} Example trajectory through state space. Without external input (control signals) the system's passive dynamics leads to a state where random brain regions are more active than others; with input the system is driven into the desired target state. \emph{(D)} Schematic illustration of all possible system-to-system transitions.}\label{figure:fig1}
\end{center}
\end{figure*}

The remainder of this report is divided into two sections. In the first, we explore a set of progressively more difficult control tasks, demonstrating that a brain region's weighted degree (strength) is highly correlated with its control energy, suggesting that regions with many, strong connections contribute disproportionately more energy than regions with few or weak connections. We also studied the effect of selectively suppressing input to select sets of brain regions, which revealed a network of compensatory interactions among brain regions. We show that the degree to which one brain region compensates for the removal of another can be predicted with the network measure ``communicability'', which measures the strength of direct and indirect pathways between network nodes.

The second section builds on the results from the first. Rather than use a pre-defined set of initial and target states, we objectively assign nodes to initial and target states so as to minimize the cost of the transitions among these states. We find that the optimal assignments implicate highly connected brain regions as ideal targets and weakly connected regions as ideal observers (i.e. they play no role in control). We also show that the energy associated with these optimal assignments is less than what would be expected given a degree-preserving random network model. Finally, we show that when the network's rich club \cite{van2011rich} is disrupted the energy increases, further suggesting that the configuration of connections among hub regions supports low-energy transitions.

\section{Mathematical Model} We studied a dynamical system in which the brain's network of white matter fiber tracts among brain regions constrained the following linear time-invariant nodal dynamics \cite{lin1974structural, gu2015controllability}:

\begin{equation}
\dot{\mathbf{x}}=\mathbf{A} \mathbf{x}(t) + \mathbf{B}_\mathcal{K} \mathbf{u}_\mathcal{K}(t).
\label{eq:eq1}
\end{equation}

\noindent Here, $\mathbf{x}(t) \in \mathbb{R}^{n \times 1}$, is the state vector whose element, $x_i(t)$, represents the state (activity level) of brain region $i$ at time $t$. The matrix $\mathbf{A} \in \mathbb{R}^{n \times n}$ is the symmetric and weighted adjacency matrix, whose element $A_{ij}$ is the number of detected tracts between regions $i$ and $j$ normalized by the sum of their volumes. The input matrix, $\mathbf{B}_\mathcal{K}$, specifies the set of control points, $\mathcal{K} = \{ k_1 \ldots k_m \}$, such that:

\begin{equation}
\mathbf{B}_\mathcal{K} = [\mathbf{e}_{k_1}, \ldots, \mathbf{e}_{k_m}]\\,
\label{eq:eq2}
\end{equation}

\noindent where $\mathbf{e}_{k_i}$ is the $i$th canonical column vector of dimension $n$. The time-varying input signals are denoted as $\mathbf{u}_\mathcal{K}(t) \in \mathbb{R}^{m \times 1}$ where $u_{k_i}(t)$ gives the input at control point $k_i$ at time $t$ (Figure~\ref{figure:fig1}).

We were interested in the control task where the system transitions from some initial state, $\mathbf{x}_0 = \mathbf{x}(t = 0)$, to some target state, $\mathbf{x}_T = \mathbf{x}(t = T)$. We solved this task using an optimal control framework, deriving the set of minimum-energy inputs, $\mathbf{u}^*_\mathcal{K}(t)$, for accomplishing this task (Materials and Methods). Each control point's energy was defined as $E_{k_i} = \int_{t = 0}^T \| u_{k_i}^*(t) \|^2 dt$ and the total energy was given by $E = \sum_{i = 1}^m E_{k_i}$.

Rather than investigate all possible transitions, we considered a limited repertoire, focusing on a set of eight states based on systems previously identified in intrinsic functional connectivity studies \cite{yeo2011organization} (Materials and Methods, Cognitive systems). We analyzed all system-to-system transitions (excluding self-transitions), resulting in 56 possible control tasks. Importantly, because our dynamical model is linear, any possible transition can be written as a linear combination of these transitions (though the resulting transition may not be optimal, in terms of minimum energy). Thus, our results are generally relevant to all transitions.

For a given control task we classified brain regions based on their states at times $t=0$ and $t=T$:

\begin{enumerate}
\item The \emph{initial} class includes all regions that were active at $t=0$ and inactive at $t=T$ (denoted as $\mathbf{x}_0$).
\item The \emph{target} class, on the other hand, included all regions that were inactive at $t=0$ but active at $t=T$ (denoted as $\mathbf{x}_T$).
\item The \emph{bulk} class included regions that inactive at both $t=0$ and $t=T$.
\end{enumerate}

In principal, there is a fourth class of regions active at both $t=0$ and $t=T$. However, given the control tasks we consider here, this case never arises. In subsequent sections we will present our results within this classification scheme.

\begin{figure*}[t]
\begin{center}
\centerline{\includegraphics[width=1\textwidth]{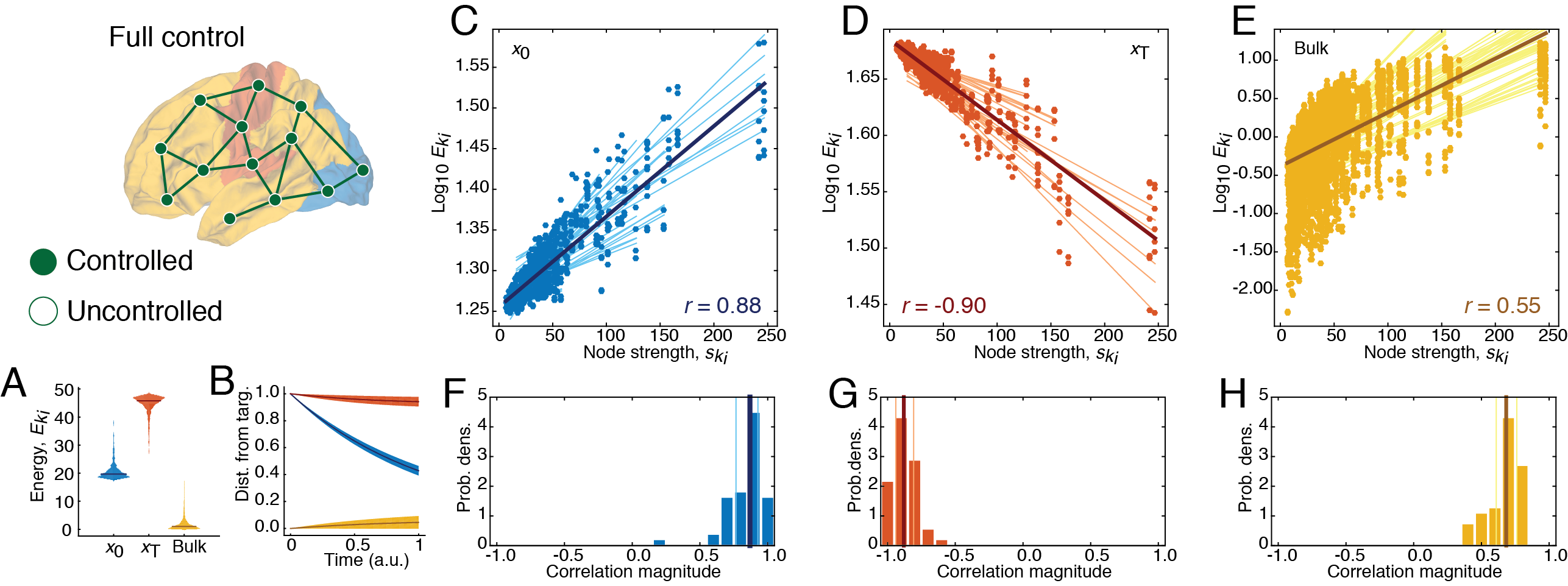}}
\caption{\textbf{Energy depends on class and strength when all regions are controlled.} \emph{(A)} Violin plot of node-level control energies aggregated across all control tasks and divided into initial ($\mathbf{x}_0$), target ($\mathbf{x}_T$), and bulk classes. \emph{(B)} Average distance of $\mathbf{x}(t)$ from target state, $\mathbf{x}_T$, under free evolution (i.e. no input signals) as a function of time and averaged within each class. \emph{(C-E)} Scatterplots of node-level control energies, $log_{10} E_{k_i}$, \emph{versus} strengths, $s_{k_i}$, (total normalized streamline counts) across all control tasks. \emph{(F-H)} Distributions of Pearson's correlation coefficients for best-fit lines shown in panels \emph{(C-E)}.}\label{figure:fig2}
\end{center}
\end{figure*}

\section{Results} 

\subsection*{Predicting energy from topology}

\begin{figure*}[t]
\begin{center}
\centerline{\includegraphics[width=1\textwidth]{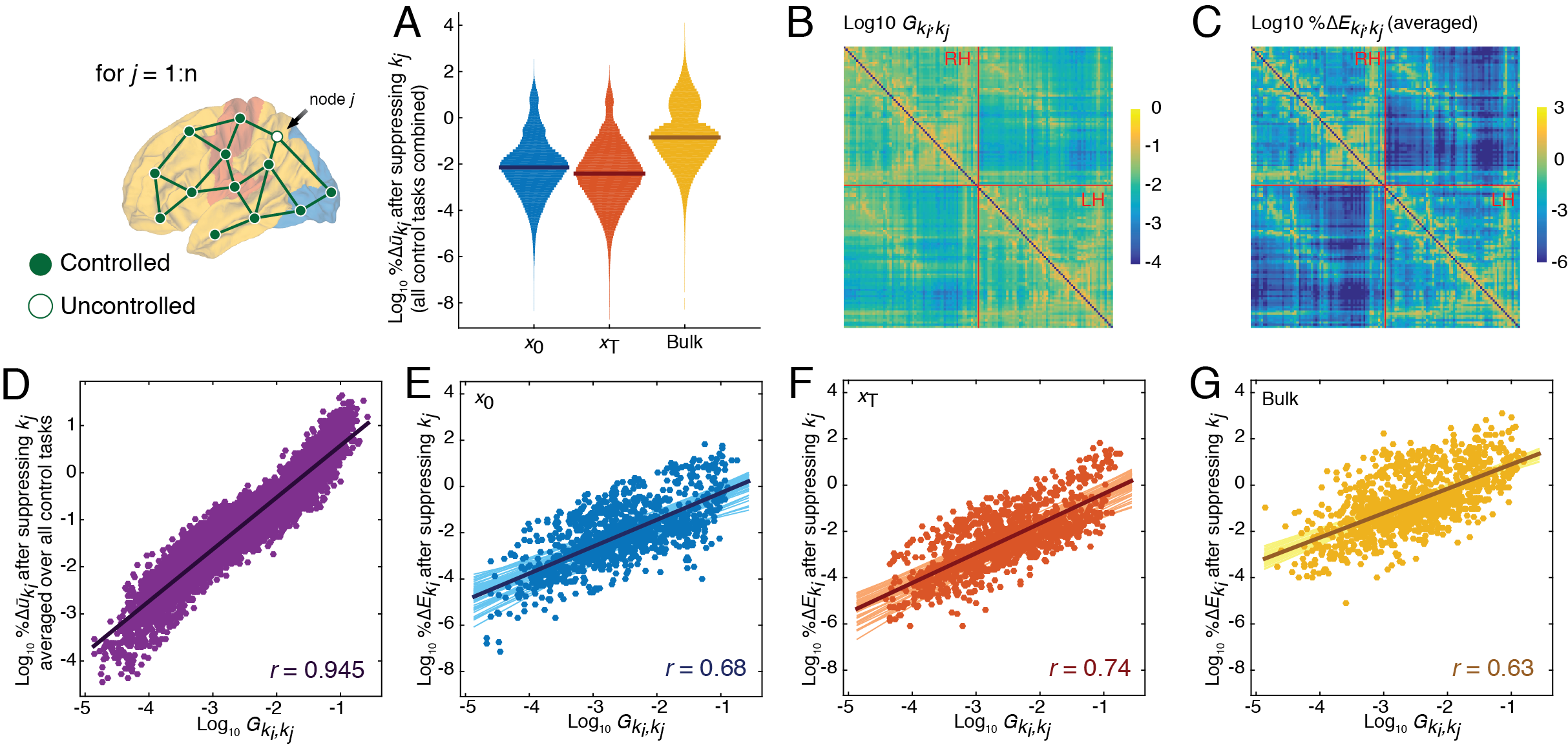}}
\caption{\textbf{Communicability predicts compensation (or change in energy) when individual regions are suppressed.} \emph{(A)} Violin plot of percent change in regions' energies ($\% \Delta E_{k_i}$) after suppressing region $k_j$ aggregated by class. \emph{(B)} Compensation network in matrix form; the weights of the compensation network are defined as the percent change in energy of region $k_i$ ($\% \Delta E_{k_i}$) after suppressing region $k_j$ averaged across all control tasks. \emph{(C)} Weighted communicability matrix, which measures the total strength of direct and indirect connections between two regions. \emph{(D)} Scatterplot showing the correlation of communicability and the compensation network weights. \emph{(E-G)} Percent change in the energy of region $k_i$ after suppressing region $k_j$. Each class is shown separately -- $\mathbf{x}_0$ are active at time $t=0$, $\mathbf{x}_T$ are active at time $t=T$, and bulk nodes are inactive at both $t=0$ and $t=T$. Matrices in panels \emph{(B)} and \emph{(C)} are shown with logarithmic scaling.}\label{figure:fig3}
\end{center}
\end{figure*}

One of the principal aims of this report was to determine what features of the brain's topology influence control energy. In this section we assessed the role of topology in determining the energy associated with each of the 56 control tasks. We began this investigation with a simple scenario in which all brain regions served as control points. We observed that $E_{target} > E_{initial} > E_{bulk}$ (one-way ANOVA comparing log-transformed energies across all control tasks and participants yielded minimum omnibus test-statistic of $F(2,126) = 83.0$, and a maximum $p$-value of $p = 1.0 \times 10^{-23}$; pairwise t-tests for $E_{target} > E_{initial}$ and $E_{initial} > E_{bulk}$ yielded minimum t-statistic of $t=8.31$ and maximum $p$-value of $p = 1.6 \times 10^{-9}$, Bonferroni adjusted) (Figure~\ref{figure:fig2}A).

To explain this ranking of energies, we need to consider the system's dynamics under free evolution -- the absence of input. In such a case, the system evolves as  $\mathbf{x}(t) = e^{\mathbf{A}t}\mathbf{x}_0$, which we obtained by solving Eq.~\ref{eq:eq1}. The vector $\mathbf{v}(t) = \mathbf{x}_T - \mathbf{x}(t)$, then, specified the distance, $v_i(t)$, of each region from its target state at time $t$. Intuitively, as distance increased, more energy was required to drive the system towards its desired configuration (Figure \ref{figure:suppDistanceVsEnergy}). Moreover, distance followed a class-specific trajectory (Figure~\ref{figure:fig2}B). At the control horizon, $t=1$, bulk regions were nearest their target state, followed by initially active regions, followed by target regions, explaining why specific classes required more or less energy.

The previous analysis demonstrated that, foremost, regional control energy depended on class. Within each class, however, we found that much of the remaining variance could be accounted for by regions' weighted degrees (\emph{strength}; $s_i = \sum_j A_{ij}$). Across participants and control tasks, the logarithm of control energy for regions assigned to initial and bulk classes was both positively correlated with the logarithm of strength (median(interquartile range) correlations of $r = 0.85 (15)$ and $r = 0.79 (15)$, respectively) while the opposite was true for target regions $r = -0.84 (13)$) (Figure~\ref{figure:fig2}C-E; Figure~\ref{suppfullControlAcrossSubjects} for summary across all participants). Collectively, these results imply that predicting a brain region's energy contribution requires both topology and contextualizing a region's role -- i.e. initial, target, bulk -- in a given control task. 

\subsection*{Simulated suppression and compensatory effects}
In the previous section we investigated a scenario in which all brain regions served as control points, making it possible to control any region's state directly. In the current section we imagined a more difficult scenario in which the system performed the same control tasks but with specific subsets of brain regions excluded from the control set. These excluded regions, then, could only be manipulated via indirect input from their neighbors -- they required help from the rest of the network to achieve their desired state. This framework, which we refer to as ``simulated suppression'', is analogous to inhibitory neuromodulation, which can be achieved externally using transcranial magnetic stimulation (TMS), in which a brain region's local activity is suppressed but still receives inputs from surrounding areas \cite{pascual2000transcranial, silvanto2008new} but could also be achieved internally via competitive or inhibitory dynamics among neuronal populations.

Using simulated suppression we investigated how the suppression of input to specific regions changed the energies of the remaining, unsuppressed, regions. We interpreted such changes as a measure of compensation. In this section, we explored a series of progressively more difficult control scenarios in which we suppressed both individual brain regions and entire classes of regions.

\subsubsection*{Suppression of individual brain regions}
In this section we suppressed individual brain regions, one at a time, and repeated the same control tasks as the previous section. We expected that with fewer control points the total control energy, $E$, would increase. We found that this was largely the case, with the greatest percent changes in energy occurring among bulk regions (Figure~\ref{figure:fig3}A).

The simulated suppression framework allowed us to calculate the extent to which brain regions engaged in compensatory relationships with one another, wherein the suppression of one region consistently resulted in increased energy of another region. We were also interested in determining to what extent these compensatory relationships could be predicted based on topological properties of the network. We hypothesized that the strength of compensatory relationships should depend upon the extent to which they were interconnected. We reasoned that even indirectly-connected regions should be able to compensate for one another provided that they were linked by many multi-step paths. This intuition can be formalized as the \emph{communicability} between two regions, a statistic that quantifies the strength of both direct and indirect pathways between node pairs \cite{estrada2008communicability}. Communicability can be thought of as the capacity for two regions to communicate with one another by pathways of all topological lengths (Figure~\ref{figure:fig3}B) (Materials and Methods, Network communicability). 

For each control task we calculated the percent change in energy of region $k_i$ after suppressing region $k_j$, which we denote as $\% \Delta E_{k_i,k_j}$ (Figure~\ref{figure:fig3}C). When averaged across all control tasks we found excellent correspondence between this measure and weighted communicability, with a correlation of $r = 0.95 (0.01)$ across participants (Figure~\ref{figure:fig3}D). The same relationship persisted (albeit attenuated) when we examined specific control tasks and sub-divided regions according to their class: for initial $r = 0.74 (0.08)$, for target $r = 0.80 (0.06)$, and for bulk $r = 0.68 (0.11)$) (Figure~\ref{figure:fig3}E-G). These results suggest that both direct and indirect connections play important roles in compensating for suppressed brain regions (Figure \ref{suppremoveSingleNodesAcrossSubjects} for summary across all participants).

\begin{figure*}[t]
\begin{center}
\centerline{\includegraphics[width=0.8\textwidth]{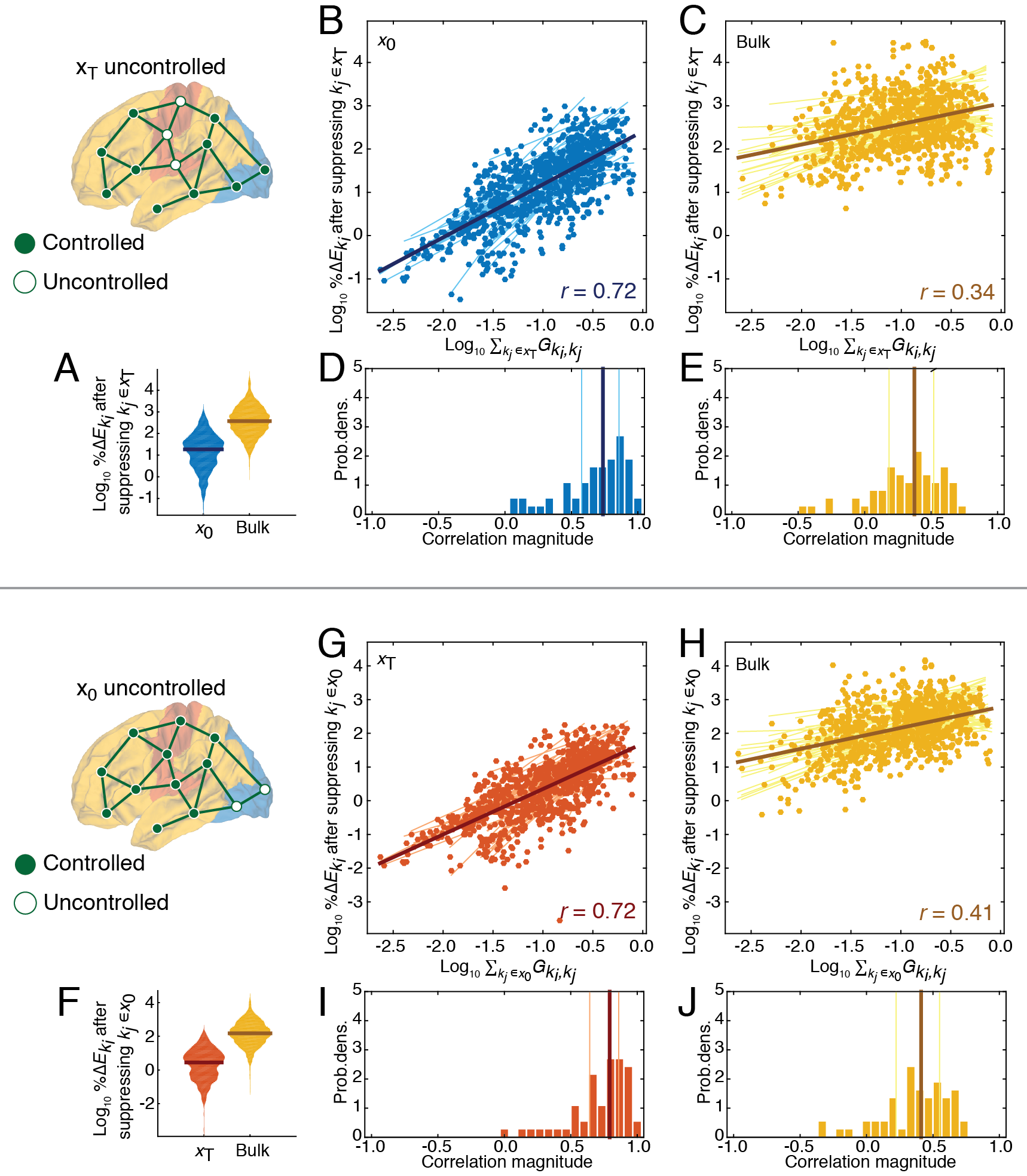}}
\caption{\textbf{Summary of initial or target class suppression.} \emph{(A)} Violin plot showing percent changes in energy $\% \Delta E_{k_i,k_j}$, for the remaining nodes after excluding all target nodes from the control set. \emph{(B,C)} Scatterplot of the percent change of region $k_i$'s energy after suppressing all target nodes against $k_i$'s communicability to all target nodes. \emph{(D,E)} Distributions of Pearson's correlation coefficients between change in energy and communicability, $\hat{r}_{G,u}$, for initial and bulk nodes. Panels \emph{(F-J)} recapitulate \emph{(A-E)} but with initial nodes excluded from the control set rather than target nodes.}\label{figure:fig4}
\end{center}
\end{figure*}

\subsubsection*{Suppressing initial \emph{or} target classes}
We extended the simulated suppression framework by simultaneously suppressing entire classes of brain regions, resulting in considerably more difficult control tasks. In particular, we suppressed all regions assigned to either initial or target classes. As in the previous section, we found that bulk regions exhibited the greatest percent change in their energies (Figures~\ref{figure:fig4}A,F). Also in agreement with the previous section, we found that the percent change in a region's energy was predicted by its total communicability to the suppressed class ($\sum_{k_j \in \mathbf{x}_0} G_{k_i,k_j}$ and $\sum_{k_j \in \mathbf{x}_T} G_{k_i,k_j}$) (Figure~\ref{figure:fig4}B,C,G,H). When the target class was suppressed, the correlation of communicability and change in energy was $r = 0.76 (0.24)$ and $r = 0.38 (0.36)$ for the remaining initial and bulk regions, respectively. When the initial class was suppressed, the correlation of communicability and change in energy was $r = 72 (0.30)$ and $r = 0.32 (0.41)$ for the remaining target and bulk regions (Figure~\ref{suppremoveEachClassAcrossSubjects} for summary across all participants).

\subsubsection*{Suppressing initial \emph{and} target classes}
\begin{figure*}[t]
\begin{center}
\centerline{\includegraphics[width=0.8\textwidth]{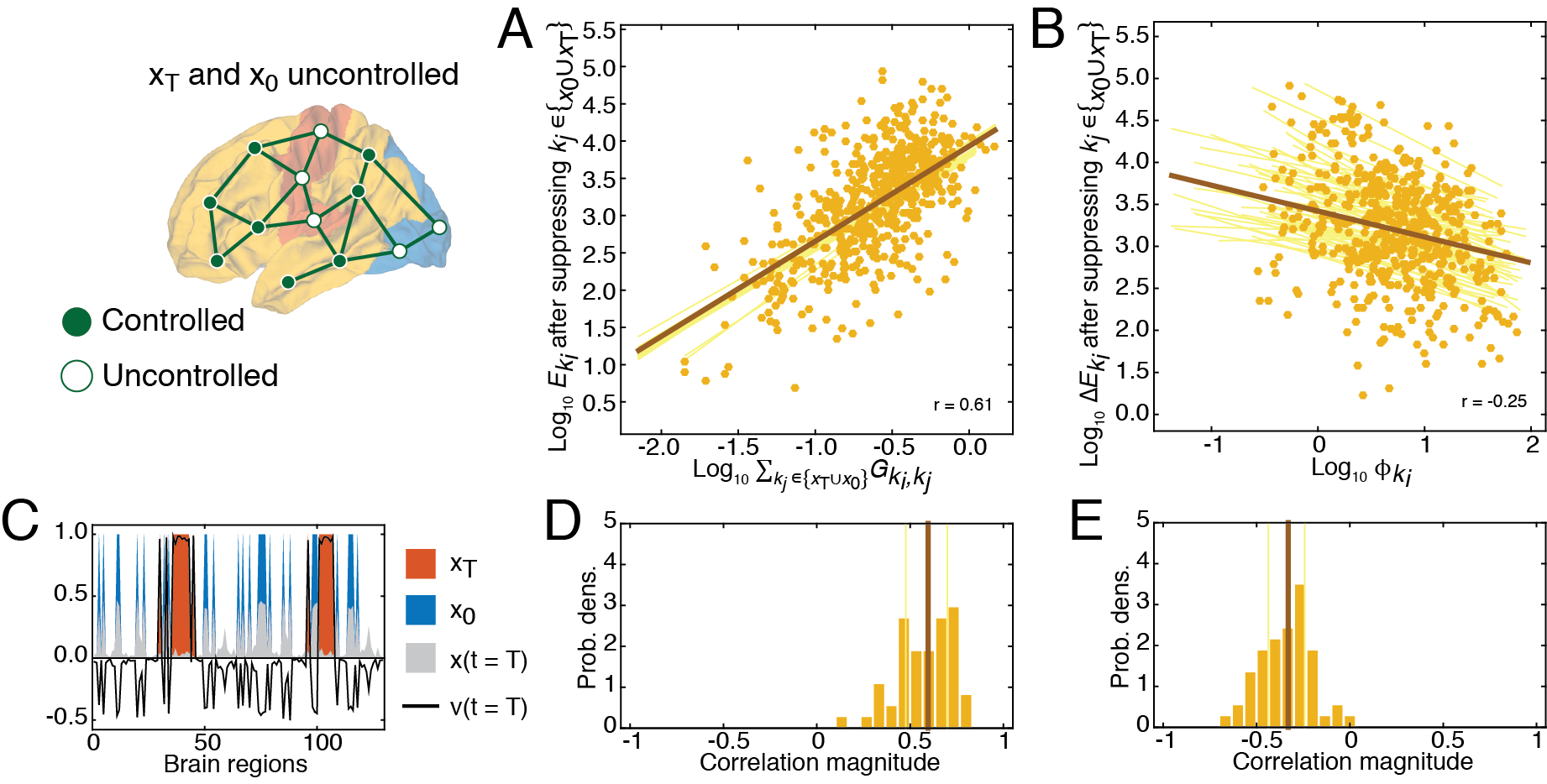}}
\caption{\textbf{Summary of initial and target class suppression} \emph{(A)} Only bulk nodes are directly controlled. Their energies, $E_{k_i}$, are strongly correlated with their communicability to the initial and target nodes, $\sum_{k_j \in \{ \mathbf{x}_0 \cup \mathbf{x}_T \}} G_{k_i,k_j}$. \emph{(B)} The percent change in total energy going from the \emph{full control} experiment to the \emph{no control} experiment was negatively correlated with $\theta_{k_i} = \sum_{j} A_{k_i,j} \| v_{j} \|$, or the weighted distance of controller $k_i$'s neighbors from their respective target states relative to their states under free evolution. \emph{(C)} An illustrative example of the components that go into calculating $\theta_{k_i}$. The blue and orange bars represent the state of the system at times $t = 0$ and $t = T$. The blue bars indicate the activation of DMN nodes, in this case, while the orange bars indicate activation of VIS nodes. The grey bars represent the state of the system at time $t = T$ under free evolution after starting in $\mathbf{x}_0 = DMN$. Finally, the black line is equal to $\mathbf{v}(t) = \mathbf{x}_T - \mathbf{x}_0$, or the element-wise difference between the orange and grey bars. \emph{(D,E)} Distributions of Pearson's correlation coefficients from the best-fit lines shown in panels \emph{(A)} and \emph{(B)}, respectively. } \label{figure:fig5}
\end{center}
\end{figure*}

Finally, we explored the consequences of simultaneously suppressing both initial and target regions, leaving only the bulk as control points. In the previous sections if suppression was applied to initial or target nodes, the remaining bulk could perform the same control task as the missing nodes. Here, however the bulk was tasked with performing two duties: simultaneously turning on and off target and initial regions, respectively. The most energetic members of the bulk were those with high levels of communicability to either initial or target regions, thereby affording them the possibility of directly and indirectly controlling the states of those classes (Figure~\ref{figure:fig5}A) (across participants, the correlation of $E_{k_i}$ and $\sum_{k_j \in \{ \mathbf{x}_0 \cup \mathbf{x}_T\}} G_{k_i,k_j}$, the communicability of region $k_i$ to both initial and target regions, was $r = 0.57 (0.25)$). These results further suggest that bulk regions -- those not actively changing their own state during a control task -- nonetheless acted as compensators when other regions become compromised.

We also were able to predict bulk regions whose energy contributions increased by the greatest amount. As noted earlier, the vector $\mathbf{v}(t) = \mathbf{x}_T - e^{\mathbf{A}t}\mathbf{x}_0$ gives the Euclidean distance in nodes' states at time $t$ with their respective target states. We hypothesized that the greater the distance a bulk region's neighbors were from their target states would be related to how much additional energy that region would have to contribute. To this end, we calculated $\mathbf{\theta} \in \mathbb{R}^{m \times 1}$ whose element $\theta_{k_i} = \sum_{j} A_{k_i,j} \| v_{j} \|$ gives the total distance of $k_i$'s neighbors from their target states weighted by the strength of $k_i$'s connection to those nodes. We found that the logarithm of $\theta_{k_i}$ and $E_{k_i}$ were robustly negatively correlated across both participants and control tasks ($r = -0.33 (0.18)$) (Figure~\ref{figure:fig5}B) (Figure~\ref{suppremoveBothClassesAcrossSubjects} for summary across all participants).

Collectively, the results reported in this section make two important points. First, bulk regions -- those not changing their state from active to inactive (or \emph{vice versa}) -- exhibited the greatest increase in energy following the suppression of other regions, suggesting that these ``bystanders'' may play important compensatory roles in the control of brain dynamics. Secondly, in demonstrating that brain region's compensatory relationships are correlated with their communicability, we implicate indirect connections as important pathways through which compensatory relationships emerge.

\subsection*{Optimal class assignments}
In the previous section we explored the relationship of control energy to network topology across a set of control tasks involving transitions to and from pre-defined states. An alternative approach, is to identify initial and target states so that the associated transition requires little energy. In this section we investigated such an approach. Specifically, we identified assignments that minimized the objective function, $\mathcal{E} = \mathbf{v}^T\mathbf{v}$, where $\mathbf{v}$ is defined the same way as before. We chose this particular objective function because it is highly correlated with control energy (Figure \ref{figure:suppDistanceVsEnergy}) and also for computational ease, as it can be calculated in a more straightforward manner than energy, which requires first deriving the optimal control signals.

We used a simulated annealing optimization algorithm to minimize $\mathcal{E}$ and generate estimates of the probability with which each brain region was assigned to initial, target, and bulk classes. Across participants we found that the probability of a region being assigned to either the target and bulk classes was highly correlated with its weighted degree ($r_{target} = 0.88 (0.03)$ and $r_{bulk} = -0.73 (0.07)$; maximum $p$-value of $p = 2.5 \times 10^{-16}$), while the nodes assigned to the initial class were not closely associated with weighted degree ($r = -0.02 (0.09)$; minimum $p$-value of $p = 0.27$). These results suggest that highly and weakly connected regions are ideal targets and bystanders, respectively.

What aspects of the network's organization determines class assignments? One possibility is that the low-energy transitions facilitated by these assignments are merely a consequence of the number and weight of connections a brain region makes -- i.e. they do not depend on the actual configuration of a network's connections. To test this hypothesis, we randomized connection placement while preserving the number of connections that each brain region makes (generating 100 random networks for each participant) and evaluated the objective function for the optimal class assignments given these networks. We observed that across all class compositions the randomized networks were associated with significantly greater energies than the observed networks (non-parametric test, max $p < 1 \times 10^{-15}$).

\subsection*{Rich club promotes low-energy transitions}

\begin{figure*}[t]
\begin{center}
\centerline{\includegraphics[width=1\textwidth]{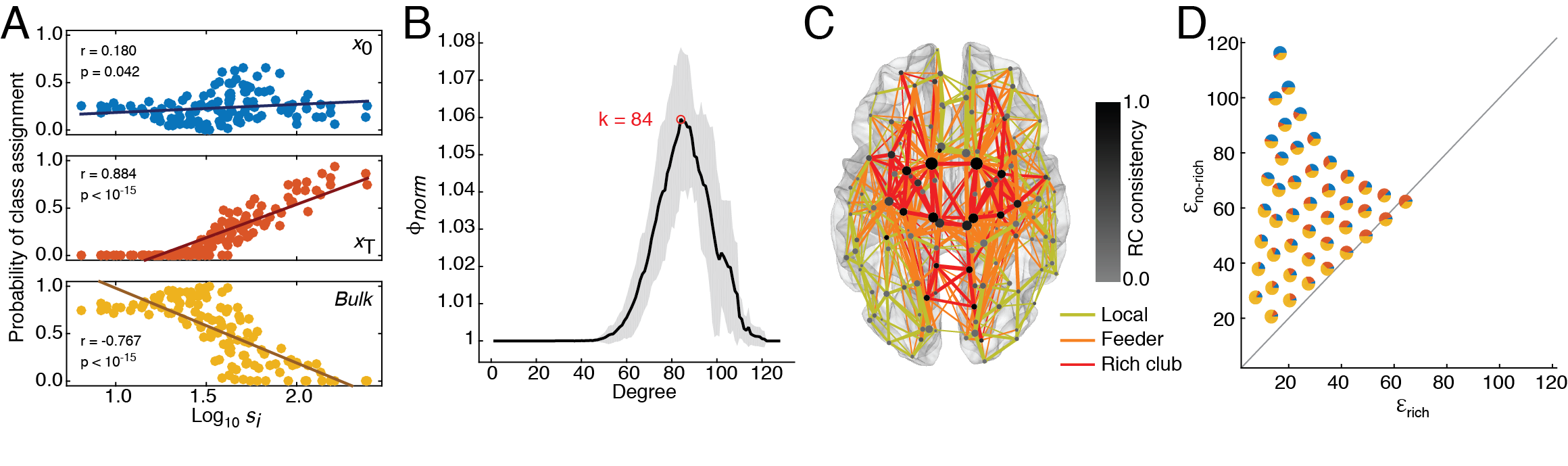}}
\caption{\textbf{Rich club detection and optimal class assignments.} \emph{(A)} Class assignment probabilities as a function of the logarithm of each brain region's strength. \emph{(B)} Normalized rich club coefficient as a function of degree. Gray envelope represents $\pm$ one standard deviation. \emph{(C)} Topographic distribution of rich club nodes and classification of edges (top 5\% for visualization) as either ``local'' (links two non-rich club regions), ``feeder'' (links a rich club region to a non-rich club region), or ``rich club'' (links a rich club region to another rich club region). The size of each node is proportional to its degree and the darkness of each point indicates the likelihood, across participants, that that node was assigned to the $k=84$ rich club. \emph{(D)} Comparison of energy with rich club intact ($\mathcal{E}_{rich}$) versus rewired rich club ($\mathcal{E}_{no-rich}$). Each point (pie chart) represents a particular composition of class assignments. The gray line indicates the ``break-even'' line -- points along the line correspond to optimal class assignments that have approximately equal average energy both with and without a rich club. Points above this line indicate that rewiring the rich club leads to increased energy associated with the optimal class assignments.} \label{figure:fig6}
\end{center}
\end{figure*}

The observation that highly-connected brain regions make good targets and poor bystanders suggests that hub regions and, perhaps, rich clubs play an important role in facilitating low-energy control \cite{van2011rich, van2013network}. Intuitively, a rich club is a collection of hubs -- highly connected, highly central regions -- that are more densely interconnected to one another than expected. This type of organization is thought to promote rapid transmission and integration of information among brain regions \cite{van2012high}. Indeed, rich club regions were more likely to be assigned to the optimal target state (Figure \ref{suppclassAssignmentsBySystemAndRC}). The previous null model, wherein all connections were randomized, tested the null hypothesis that structureless networks could produce comparable levels of energy. To test whether a network's rich club influences energy, we require a more subtle and specific null model. Moreover, a network's rich club is a pseudo-continuous structure and can be defined at multiple resolutions. Our focus was on the neighborhood of binary rich clubs identified at $k=84$ -- i.e. all brain regions included in the rich club must have degree of at least 84 -- which corresponds to the maximum normalized rich club coefficient obtained across all participants (Figure \ref{figure:fig6}A). At this level, the most consistent rich club regions across participants included subcortical regions thalamus, caudate, putamem and hippocampus as well as precuneus, isthmus cingulate, posterior cingulate, lateral orbito-frontal, and insular cortex. These regions are in close agreement with previously-described rich clubs and hubs \cite{hagmann2008mapping,van2011rich, van2013network, betzel2014changes, crossley2014hubs}.

We implemented a null model where we rewired only connections among rich club members (while preserving degree) so that the density of connections among rich club members was as low as possible. We observed that as long as eight brain regions (the smallest increment that we considered) were assigned as initially-active regions, then rewiring connections to dissolve the rich club alway yields increased energy (non-parametric test, max $p < 1 \times 10^{-15}$), suggesting that the brain's rich club, specifically, supports low-energy transitions from a diverse set of initial states to a target state of high-strength hub regions (we verify that this result holds for rich clubs defined at $k=80$ to $k=88$; Figure \ref{supprichClubAtDifferentK}).

\section*{Discussion}
In this report we used network control theory to investigate the role of the brain's anatomical network in supporting transitions among different brain systems. In the first section we focused on transitions among a pre-defined set of brain systems and demonstrated, in agreement with earlier work, that brain regions with many strong connections were associated with increased energies. We further demonstrated that when the control set is perturbed, by suppressing control of both individual brain regions and entire classes, the remaining regions compensate for the loss by increasing their own energies. Moreover, we showed that the percent change in a controller's energy could be predicted by its communicability to the suppressed regions, highlighting the role of indirect communication paths. In the second section, we sought to objectively identify initial and target states that could be transitioned to and from for little energy. We found that the optimal initial states were diverse while the probability that a region was among the target state was highly correlated with its weighted degree. We showed that transitions among these optimal class assignments were, in part, facilitated by the brain's rich club; when connections among rich club regions were rewired, the energy associated with such transitions consistently increased.

\subsection*{From descriptive to predictive network models.} The study of networked neural systems has advanced rapidly in the past decade. While early analyses focused on the topological properties of SC networks such as their small-worldness \cite{iturria2008studying, gong2009mapping} or the presence of hubs, rich clubs, and modules \cite{hagmann2008mapping, van2011rich}, the focus of recent work has shifted from static descriptions to dynamical systems models, making it possible to investigate how network topology shapes passive dynamics (i.e. no inputs) \cite{breakspear2010generative, haimovici2013brain, abdelnour2014network, misic2015cooperative}. The natural extension of these and other recent studies is to incorporate exogenous input into the dynamical model. Such an extension makes it possible to begin addressing questions related to the control of the brain. At the level of large-scale human brain networks, these theoretical questions are only now beginning to be addressed \cite{gu2015controllability, muldoon2016stimulation}, though the utility of this approach is obvious, showing promise in stimulation-based treatment of epilepsy \cite{taylor2015optimal}. These first studies contributed statistics for characterizing the extent to which brain regions contribute to making the entire state space accessible for a network -- i.e. rendering it controllable. The entire state space, however, likely contains states that, for one reason or another, should actively be avoided by the system. In this present study we sought to characterize the energy contributions of brain regions based on transitions among a limited state space populated by seemingly neurophysiologically plausible states.

\subsection*{Structural predictors of the ease or difficulty of control} The aim of this report was to shed light on the features of a network's topology that contribute to making a control task easier or difficult. Toward this end we made a number of contributions and novel observations. First, we presented a classification system of nodes for studying specific control tasks. We showed that for the control tasks we investigated, a node's control energy is highly correlated with its strength. The nature of this relationship, however, depends critically on whether a node is classified as part of the initial, target, or bulk set -- the energies of nodes that are initially active and later inactive (initial) as well as nodes that, at the boundary conditions of $t=0$ and $t=T$, are inactive (bulk) maintain strong positive relationships with node strength; the opposite is true for target nodes, which are negatively correlated with node strength.

Under this framework, we explored a series of progressively more difficult control tasks in which we suppressed specific subsets of nodes. This set of experiments highlighted the brain's compensation network -- a network whose edge weights represent the percent change in a node's energy when other nodes are excluded from the control set. We went on to show that \emph{communicability} between two nodes was highly correlated with their compensation weight, which is interesting for several reasons. Communicability measures the strength of direct and indirect connections between two nodes  \cite{estrada2012physics}, which suggests a possible functional role for multi-step pathways in the human brain. Namely, that when a brain region's capacity for control is compromised -- e.g. acute ischemic stroke or electro-magnetic simulation -- the regions that ``pick up the slack'' and take on expanded control roles include, as one might anticipate, those with direct connections to the compromised region, but also those with many indirect and potentially long-distance connections. This observation serves as a potential mechanistic account of diaschiatic phenomena, where the effect of a focal lesion on brain function is observed some distance from the lesion site \cite{grefkes2011reorganization, carrera2014diaschisis, crofts2011network}. Additionally, this observation is in line with recent work showing that including multi-step pathways in predictive models of FC leads to decreased error rates and improved predictions of resting state functional connectivity \cite{goni2014resting, becker2015accurately}.

Interestingly, we observed that bulk regions exhibited the greatest percent increase in their energy. This suggests that regions not directly involved in a particular control task actually play a disproportionately greater compensatory role than those directly involved -- i.e. initial and target regions. Intuitively, these results may provide a context to understand cognitive dysfunction observed in neurological conditions that involve region damage or loss, such as traumatic brain injury or Alzheimer's disease. From a control theoretic perspective, as brain regions suffer damage, the increased burden to bulk nodes for multiple control tasks may interfere with one another. Bulk regions may become sites of processing interference due to competition between new compensatory roles. This increased competition between control tasks could have implications for the consequences of brain injury and neurodegeneration over longer timescales \cite{sivanandam2012traumatic}. In particular, increased processing burdens may result in later deleterious effects on regions that assume a disproportionate share of compensatory burdens \cite{stam2014modern}, resulting in a cascade of later failures across the brain.

Regional suppression is also relevant to models of conflict processing and the cognitive effort expended in doing so \cite{kurzban2013opportunity}. In such models, computational executive control mechanisms can be deployed for a limited number of simultaneous tasks, resulting in opportunity costs. With increased opportunity costs comes the phenomenology of effort (motivation, fatigue, boredom), which, in turn, alters the control mechanisms. In the context of our control theoretic view of task control, as node suppression introduces potential conflicts between control tasks among bulk regions, we might expect increases in perceived effort, difficulty managing sustained tasks or task switches, and negative consequences for task performance in a variety of disorders and tasks with control demands. Future studies could test this prediction by mapping the cognitive difficulties and perceived effort that arise from conflicting control demands following real region damage or following non-invasive temporary suppression. Specifically, we would expect that cognitive performance would be low and perceived effort high in general in individuals where a relatively small set of bulk regions assume a relatively high set of compensatory roles across control tasks. Thus, our control theoretic approach may provide a practical means to quantify effort and processing conflicts in health and disease.

\subsection*{The rich club as control backbone}
Whereas we have approached control from an engineering perspective, in cognitive neuroscience the term control refers to a set of processes, which include memory and attentional systems, for guiding behavior towards a particular goal \cite{schneider1977controlled, norman1986attention}. Control is thought to be instantiated via an anatomical substrate consisting of cortical \cite{miller2001integrative} and sub-cortical areas \cite{miller2001integrative} acting largely through inhibitory mechanisms. An important question concerns how this neuro-mechanistic account of control interfaces with current theories on the role of brain connectivity in normative brain function and to what extent it is related to the engineering-focused approach adopted here. In the second section we identified brain regions that act, in an objective sense as optimal initial, target, and bulk classes for control. We showed that the probability of a region being classified as target or bulk was closely related to its strength, with high and low strength regions more likely to be classified as targets and bulk, respectively. The initial class, on the other hand, had a more diverse constituency and was not obviously related to strength. We went on to show that the energies associated with these class assignments were much lower than those obtained from degree-preserved random networks, indicating that the class assignments were driven by some non-trivial aspect of the network's topology. We further showed that when the brain's rich club was dissolved the energy associated with these optimal assignments increased, suggesting that the rich club contributes in facilitating low-energy transitions from among a diverse set of initial states to target states composed of high-strength, high-degree rich club regions. 

The rich club is typically thought of as an integrative backbone that allows hub regions to communicate with one another, facilitating rapid communication and transmission of signals across the entire brain \cite{van2012high, senden2014rich, collin2014structural}. Despite this supposition, there is some controversy as to the rich club's precise role in network communication, with some indication that rich club hubs are primarily drivers of network dynamics -- acting in a top-down fashion to influence different sub-systems \cite{harriger2012rich, towlson2013rich, van2013anatomical}. On the other hand, rich club regions are highly connected to the rest of the network and, by this virtue, are not only in a position to influence their neighbors, but also to be influenced \emph{by} their neighbors. Indeed, computational models suggest that rich clubs adopt stable, regular behavior as a consequence of receiving input from a multitude of sources \cite{gollo2015dwelling}. Our results agree with this latter account in some respects. We observed that high-degree, rich club regions are best suited for roles as targets that can be transitioned into from a diverse set of initial states. Part of why rich club regions are so successful in this respect is by virtue of their many connections -- no matter where input is injected into the system, there is a high probability that the signal will propagate from its source to activate regions in the rich club, suggesting a possible explanation for why rich club regions, which overlap considerably with the brain's default mode network, tend to be active in the default mode of resting state function \cite{van2013anatomical}. More relevant to the present discussion, this also suggests that under certain conditions the rich club (or at least highly connected regions) may be in a better position to be indirectly controlled than weakly connected regions on the network periphery.

\subsection*{Methodological considerations}

Several important methodological considerations are pertinent to this work. First, we relied on diffusion spectrum imaging and tractography to infer the presence of large-scale fiber tracts in participants' brains. These methodologies are imperfect and can detect spurious tracts or fail to detect existing tracts \cite{thomas2014anatomical, reveley2015superficial}. However, at present there exist no other non-invasive methods for reconstructing human structural connectivity networks. Future work will likely address these shortcomings by introducing improved tractography algorithms and imaging techniques \cite{pestilli2014evaluation}.

Another important consideration is our use of a linear dynamical model despite the fact that, by most accounts, brain activity is fundamentally non-linear \cite{deco2011emerging}. Our justification for using such dynamics is twofold. First, the emphasis of this paper is on the role that the brain's structural connectivity network plays in control. While the form of dynamics certainly contributes to making a system controllable or not, we focus primarily on the contribution of the network's topology. Secondly, a linear dynamics is in line with other papers that investigate control theory with an emphasis on topology \cite{liu2011controllability, muller2011few, yan2012controlling}. Moreover, describing non-linear systems in terms of a linear approximation in the neighborhood of its equilibrium points is common \cite{luenberger1979introduction}, and makes it possible to apply linear control to otherwise intractable systems.

A final limitation is the form of the communicability measure, which (when applied to a binary networks) weighs longer paths exponentially less than shorter paths. While this standard form appears sufficient for our purposes here, alternative weighting schemes may provide additional insight in the role of multi-step paths on network control.

\subsection*{Conclusion}

Understanding how control occurs in the brain, and how we can use external interventions to affect that control, has broad implications across the cognitive and clinical neurosciences. By examining control strategies implemented in finite time and with limited energy, we were able to uncover fundamental principles of brain structure that impact the ease or difficulty of control tasks informed by the systems known to perform diverse cognitive functions. It is intuitively plausible that these principles may be altered by psychiatric disease and neurological disorders, via a change in underlying structural connectivity. In future, it will be interesting to understand how individual differences in brain structure affect individual differences in the natural implementation of control (e.g., cognitive control) and resulting behavior in executive domains. Moreover, it will be interesting to understand how these individual differences might also directly affect susceptibility and response to external interventions via invasive or non-invasive neuromodulation. Such an understanding would provide critical groundwork for personalized medicine.

\section*{Materials and Methods}

\subsection*{Data acquisition and processing}
Diffusion spectrum images (DSI) were acquired for a total of 30 subjects along with a T1-weighted anatomical scan at each scanning session. We followed a parallel strategy for data acquisition and construction of streamline adjacency matrices as in previous work \cite{gu2015controllability}. DSI scans sampled 257 directions using a Q5 half-shell acquisition scheme with a maximum $b$-value of 5,000 and an isotropic voxel size of 2.4 mm. We utilized an axial acquisition with the following parameters: repetition time (TR) = 5 s, echo time (TE)= 138 ms, 52 slices, field of view (FoV) (231, 231, 125 mm). All participants volunteered with informed consent in accordance with the Institutional Review Board/Human Subjects Committee, University of Pennsylvania.

DSI data were reconstructed in DSI Studio (www.dsi-studio.labsolver.org) using $q$-space diffeomorphic reconstruction (QSDR)\cite{yeh2011estimation}. QSDR first reconstructs diffusion-weighted images in native space and computes the quantitative anisotropy (QA) in each voxel. These QA values are used to warp the brain to a template QA volume in Montreal Neurological Institute (MNI) space using the statistical parametric mapping (SPM) nonlinear registration algorithm. Once in MNI space, spin density functions were again reconstructed with a mean diffusion distance of 1.25 mm using three fiber orientations per voxel. Fiber tracking was performed in DSI studio with an angular cutoff of 55$^\circ$, step size of 1.0 mm, minimum length of 10 mm, spin density function smoothing of 0.0, maximum length of 400 mm and a QA threshold determined by DWI signal in the colony-stimulating factor. Deterministic fiber tracking using a modified FACT algorithm was performed until 1,000,000 streamlines were reconstructed for each individual.

Anatomical scans were segmented using FreeSurfer59 and parcellated using the connectome mapping toolkit \cite{cammoun2012mapping}. A parcellation scheme including $n=129$ regions was registered to the B0 volume from each subject's DSI data. The B0 to MNI voxel mapping produced via QSDR was used to map region labels from native space to MNI coordinates. To extend region labels through the grey-white matter interface, the atlas was dilated by 4 mm \cite{cieslak2014local}. Dilation was accomplished by filling non-labelled voxels with the statistical mode of their neighbors' labels. In the event of a tie, one of the modes was arbitrarily selected. Each streamline was labelled according to its terminal region pair. From these data, we constructed a structural connectivity matrix, $\mathbf{A}$ whose element $A_{ij}$  represented the number of streamlines connecting different regions, divided by the sum of volumes for regions $i$ and $j$.

\subsection*{Cognitive systems}
The human brain can also be studied as a network of functional connections. Functional connectivity networks are modular, which means that they can be partitioned into non-overlapping sub-systems \cite{yeo2011organization, power2011functional, sporns2015modular}. These sub-systems are referred to as \emph{intrinsic connectivity networks} (ICNs) and have distinct cognitive and behavioral fingerprints \cite{smith2009correspondence, crossley2013cognitive}. The ICN definition used here was based on the canonical systems defined in \cite{yeo2011organization}, and included default mode (DMN), control (CONT), dorsal attention (DAN), saliency/ventral attention (SAL/VAN), somatomotor (SM), visual (VIS), limbic (LIM), and sub-cortical (SUB) systems. In order to assign each region of interest to a single system, we mapped both atlases to a common surface template (\emph{fsaverage}) and calculated the overlap (number of common vertices) of each region of interest with each of the seven ICNs. A region's ICN assignment was defined as the system with which it overlapped to the greatest extent (Figure \ref{figure:suppBrainSystems}).

\subsection*{Control tasks}
We considered transitions from an initial state, $\mathbf{x}_0$, to a target state, $\mathbf{x}_T$, where $T=1$ is the control horizon. We chose this particular horizon in order to ensure that the system has time to evolve from its initial state (and thereby require some corrective input signals) but also not to be trivially long. Initial and target states were selected correspond to specific cognitive systems. For example, one possible control task was to start in a state where the default mode network (DMN) was maximally active, and to transition to a state where the visual system (VIS) becomes maximally active. Intuitively, such a transition might correspond to the presentation of a visual stimulus at rest, eliciting activation of visual cortex while suppressing activation of the default mode system. In this context, the control question one asks is which nodes play a role in the minimum energy trajectory between these states. We modeled this control task by starting DMN regions in an active state at $t=0$:

\[
    x_i (t = 0) = 
\begin{cases}
    0 ,& \text{if } i \notin \text{DMN}\\
    1,& \text{if } i \in \text{DMN}.
\end{cases}
\]

\noindent Similarly, when $t=T$, only visual regions were in an active state:

\[
    x_i (t = T) = 
\begin{cases}
    0 ,& \text{if } i \notin \text{VIS}\\
    1,& \text{if } i \in \text{VIS}.
\end{cases}
\]

\noindent Given these boundary conditions, we calculated the optimal inputs, $\mathbf{u}_\mathcal{K}^*(t)$ to effect the transition from specified initial state to specified target state. In general, state vectors at other times can take on any real value. Though we considered this limited set of states, the linear model of brain dynamics means that any transition can be written as a linear combination of the transitions we studied here.

\subsection*{Network communicability}
The adjacency matrix, $\mathbf{A}$, encodes a network's direct connections. In addition, communication between pairs of nodes can also take advantage of indirect connections. To quantify the extent to which nodes are connected indirectly, one can calculate the communicability matrix \cite{estrada2008communicability}, $\mathbf{G} \in \mathbb{R}^{n \times n}$, whose element $G_{ij} = \sum_{k=0}^\infty (\frac{\mathbf{A}^k}{k!})_{ij} = (e^\mathbf{A})_{ij}$. The element $G_{ij}$, then, represents the weighted sum of walks of all lengths. The $k!$ in the denominator means that longer walks contribute disproportionately less compared to shorter walks. Communicability has been generalized to weighted networks, such as those considered here \cite{crofts2009weighted}. Specifically, $G_{ij}^{w} = e^{\mathbf{A}'}_{ij}$, where $\mathbf{A}' = \mathbf{D}^{-\frac{1}{2}} \mathbf{A} \mathbf{D}^{-\frac{1}{2}}$ and $\mathbf{D} \in \mathbb{R}^{n \times n}$ is the square matrix whose diagonal elements $D_{ii} = \sum_j A_{ij}$. Communicability is also related to the linear dynamics we study here. Solving (\ref{eq:eq1}) for the special case where there is no input, the evolution of the system is described by the matrix exponential of the connectivity matrix.

\subsection*{Optimal class assignments and simulated annealing}
One of the aims of this paper was to identify objectively optimal initial and target class assignments. Because we did not know, \emph{a priori}, the number of nodes that should be assigned to either class, we treated the number of nodes assigned to each class as a free parameter. Our investigation of this parameter space involved fixing the number of bulk nodes (48, 56, 64, 72, 80, 88, 96, 104) and, of the remaining nodes, splitting them among initial and target classes in different proportions. For example, with 56 bulk nodes, the remaining 72 nodes were split among initial/target class in proportions of 8/64, 16/56, 24/48, 36/36, 48/24, 56/16, and 64/8. This division lead to 44 distinct class compositions. For each composition we used a simulated annealing algorithm to gradually reduce $\mathcal{E}$ and obtain estimates of the optimal class assignments. We ran the annealing algorithm 50 times for each composition. Each annealing run consisted of 100 stages. During each stage, the temperature, $\tau$, was fixed at a particular value and we performed 5000 iterations, where an iteration consisted of selecting two nodes of opposite classes at random and swapping their class assignments, after which we recalculated $\mathcal{E}'$. If $\mathcal{E}' < \mathcal{E}$ or with probability $Pr = e^{\frac{\mathcal{E}' - \mathcal{E}}{\tau}}$ we retained the new assignments for the next iteration. For all runs we used a starting temperature of $\tau_0 = 1$ which we decreased by $0.925$ with each stage so that by the final stage the temperature was $\tau \approx 4.11 \times 10^{-4}$. During each annealing run, we retained all unique class assignments. After the algorithm terminated, we retained the 100 unique assignments with lowest energies for further analysis.

\subsection*{Rich club detection} A rich club refers to a subset of nodes, all of degree $\ge k$, that are densely connected to one another \cite{colizza2006detecting, mcauley2007rich}. The density of any such subset is summarized by its rich club coefficient: $\phi(k) = \frac{2E_{>k}}{N_{>k}(N_{>k} - 1)}$. We calculated $\phi(k)$ for all possible values of $k$ across all participants and subsequently normalized rich club coefficients by comparing them against the distribution of coefficients obtained from an ensemble of randomized networks (preserved degree sequence) \cite{maslov2002specificity}. It is sometimes the case that there is no single rich club -- there may be many statistically significant rich clubs spanning a range of $k$ \cite{van2011rich}. For ease of description, however, we focus on the rich club detected at $k=84$, which corresponds to the greatest value of $\phi_{norm}$ averaged across participants (Figure~\ref{figure:fig6}A). We explore the consistency of this rich club in Figure~\ref{supprichClubAtDifferentK}.

~\\
\noindent \textbf{Acknowledgements.} This work was supported from the John D. and Catherine T. MacArthur Foundation, the Alfred P. Sloan Foundation, the Army Research Laboratory and the Army Research Office through contract numbers W911NF-10-2-0022 and W911NF-14-1-0679, the National Institute of Mental Health (2-R01-DC-009209-11), the National Institute of Child Health and Human Development (1R01HD086888-01), the Office of Naval Research, and the National Science Foundation (BCS-1441502 and BCS-1430087). FP acknowledges support from BCS-1430280. JDM acknowledges support from the Office of the Director at the National Institutes of Health (1DP5OD021352-01). 

\appendix*
\section{Optimal control}
There can be an infinite number of inputs, $\mathbf{u}_\mathcal{K}$, for driving a system from $\mathbf{x}_0$ to $\mathbf{x}_T$. We wish to find the inputs corresponding to the following minimization problem:

\begin{equation}
\min_\mathbf{u} \int_0^T (\mathbf{x}_T - \mathbf{x})^T(\mathbf{x}_T - \mathbf{x}) + \rho \mathbf{u}_\mathcal{K}^T \mathbf{u}_\mathcal{K}.
\end{equation}

\noindent Here, $\rho \in \mathbb{R}_{> 0}$ is a free parameter that scales the relative importance of the first term to the second term in the integral. We set $\rho = 100$. To identify the optimal inputs, $\mathbf{u}_\mathcal{K}$, we define the Hamiltonian:


\begin{equation}
H(\mathbf{p},\mathbf{x},\mathbf{u},t) = \mathbf{x}^T \mathbf{x} + \rho\mathbf{u}_\mathcal{K}^T \mathbf{u}_\mathcal{K} + \mathbf{p}(\mathbf{A} \mathbf{x} + \mathbf{B}_\mathcal{K} \mathbf{u}).
\end{equation}

\noindent In this expression $\mathbf{A}$ is a scaled version of the weighted connectivity matrix. Specifically, we divide the original matrix by its largest eigenvalue and subtract 1 from all diagonal elements. This effectively ensures that all eigenvalues are less than zero and renders the system stable. From the Pontryagin minimization principle, if $\mathbf{u}^*_{\mathcal{K}}$ is an optimal solution to the minimization problem with corresponding trajectory, $\mathbf{x}^*$, then there exists $\mathbf{p}^*$ such that:

\begin{equation}
\frac{\partial H}{\partial \mathbf{x}} = -2 (\mathbf{x}_T - \mathbf{x}^*) + \mathbf{A}^T \mathbf{p}^* = \dot{\mathbf{p}}^*
\end{equation}

\begin{equation}
\frac{\partial H}{\partial \mathbf{x}} = 2 \rho \mathbf{u}_\mathcal{K}^* + \mathbf{B}_\mathcal{K}^T \mathbf{p}^* = 0.
\end{equation}

\noindent This set of equations reduces to:

\begin{equation}
\begin{bmatrix}
\dot{\mathbf{x}}^* \\
\dot{\mathbf{p}}^*
\end{bmatrix}
=
\begin{bmatrix}
\mathbf{A} & -(2 \rho)^{-1} \mathbf{B} \mathbf{B}^T \\
-2 \mathbf{I} & -\mathbf{A}^T
\end{bmatrix}
\begin{bmatrix}
\mathbf{x}^* \\
\mathbf{p}^*
\end{bmatrix}
+
\begin{bmatrix}
\mathbf{0} \\
\mathbf{I}
\end{bmatrix}
2 \mathbf{x}_T
\end{equation}.

\noindent If we denote:

\begin{equation}
\tilde{\mathbf{A}}
=
\begin{bmatrix}
\mathbf{A} & -(2 \rho)^{-1} \mathbf{B} \mathbf{B}^T \\
-2 \mathbf{I} & -\mathbf{A}^T
\end{bmatrix}
\end{equation}

\begin{equation}
\tilde{\mathbf{x}}
=
\begin{bmatrix}
\mathbf{x}* \\
\mathbf{p}*
\end{bmatrix}
\end{equation}

\begin{equation}
\tilde{\mathbf{b}}
=
\begin{bmatrix}
\mathbf{0} \\
\mathbf{I}
\end{bmatrix}
2 \mathbf{x}_T
\end{equation}

\noindent then we can then write the reduced equation as:

\begin{equation}
\dot{\tilde{\mathbf{x}}} = \tilde{\mathbf{A}} \tilde{\mathbf{x}} + \tilde{\mathbf{b}}
\end{equation}

\noindent which we can solve as:

\begin{equation}
\tilde{\mathbf{x}}(t) = e^{\tilde{\mathbf{A}}t}\tilde{\mathbf{x}}(0) + \int_0^t [ e^{\mathbf{A}(t - \tau)}\tilde{\mathbf{b}} ] d\tau
\end{equation}

\noindent or, alternatively

\begin{equation}
\tilde{\mathbf{x}}(t) = e^{\tilde{\mathbf{A}}t}\tilde{\mathbf{x}}(0) + \mathbf{A}^{-1}(e^{\mathbf{A}t} - \mathbf{I}) \tilde{\mathbf{b}}.
\end{equation}

\noindent Then, substituting $t=T$, we arrive at:

\begin{equation}
\tilde{\mathbf{x}}(T) = e^{\tilde{\mathbf{A}}T}\tilde{\mathbf{x}}(0) + \mathbf{A}^{-1}(e^{\mathbf{A}T} - \mathbf{I}) \tilde{\mathbf{b}}.
\end{equation}

\noindent Let

\begin{equation}
\mathbf{c} = \mathbf{A}^{-1}(e^{\mathbf{A}T} - \mathbf{I}) \tilde{\mathbf{b}}.
\end{equation}

\noindent We can then write:

\begin{equation}
\begin{bmatrix}
\mathbf{x}^*(T) \\
\mathbf{p}^*(T)
\end{bmatrix}
=
\begin{bmatrix}
\mathbf{E}_{11} & \mathbf{E}_{12} \\
\mathbf{E}_{21} & \mathbf{E}_{22} \\
\end{bmatrix}
\begin{bmatrix}
\mathbf{x}^*(0) \\
\mathbf{p}^*(0)
\end{bmatrix}
+
\begin{bmatrix}
\mathbf{c}_1 \\
\mathbf{c}_2
\end{bmatrix}
\end{equation}

\noindent Rewriting this, we get:

\begin{equation}
\mathbf{x}^*(T) = \mathbf{E}_{11} \mathbf{x}^*(0) + \mathbf{E}_{12} \mathbf{p}^*(0) + \mathbf{c}_1
\end{equation}

\noindent which can be rearranged to write:

\begin{equation}
\mathbf{p}^*(0) = \mathbf{E}_{12}^{-1} [ \mathbf{x}^*(T) - \mathbf{E}_{11} \mathbf{x}^*(0) - \mathbf{c}_1 ]
\end{equation}

\noindent Given $\mathbf{p}^*(0)$ and $\mathbf{x}_0$, we can then integrate $\tilde{\mathbf{x}}$ forward, thereby obtaining $\mathbf{x}_T$ from which we subsequently obtain the optimal inputs, $\mathbf{u}^*_\mathcal{K}$.

\bibliography{../biblio}

\clearpage

\beginsupplement

\begin{figure}
\begin{center}
\centerline{\includegraphics[width=0.5\textwidth]{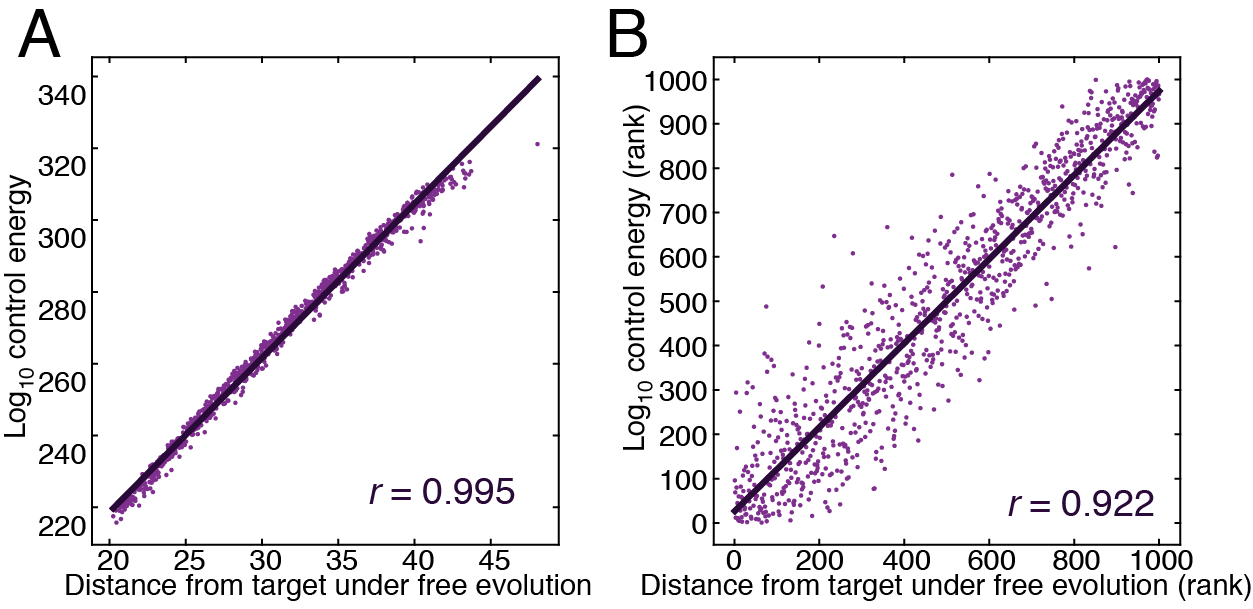}}
\caption{\textbf{Distance from target is correlated with control energy.} We generated 1000 random initial and target states and calculated both the total control energy, $E$, associated with the control task as well as the distance that the system would be from its target state given that it started in the initial state and evolved with no exogenous input, $\| v\|$. In calculating $E$ we assumed that all nodes were directly controlled. \emph{(A)} Scatterplot of the raw $\log_{10} E$ against $\| v \|$. \emph{(B)} Scatterplot of ranked $\log_{10} E$ against $\| v \|$.} \label{figure:suppDistanceVsEnergy}
\end{center}
\end{figure}

\begin{figure}
\begin{center}
\centerline{\includegraphics[width=0.5\textwidth]{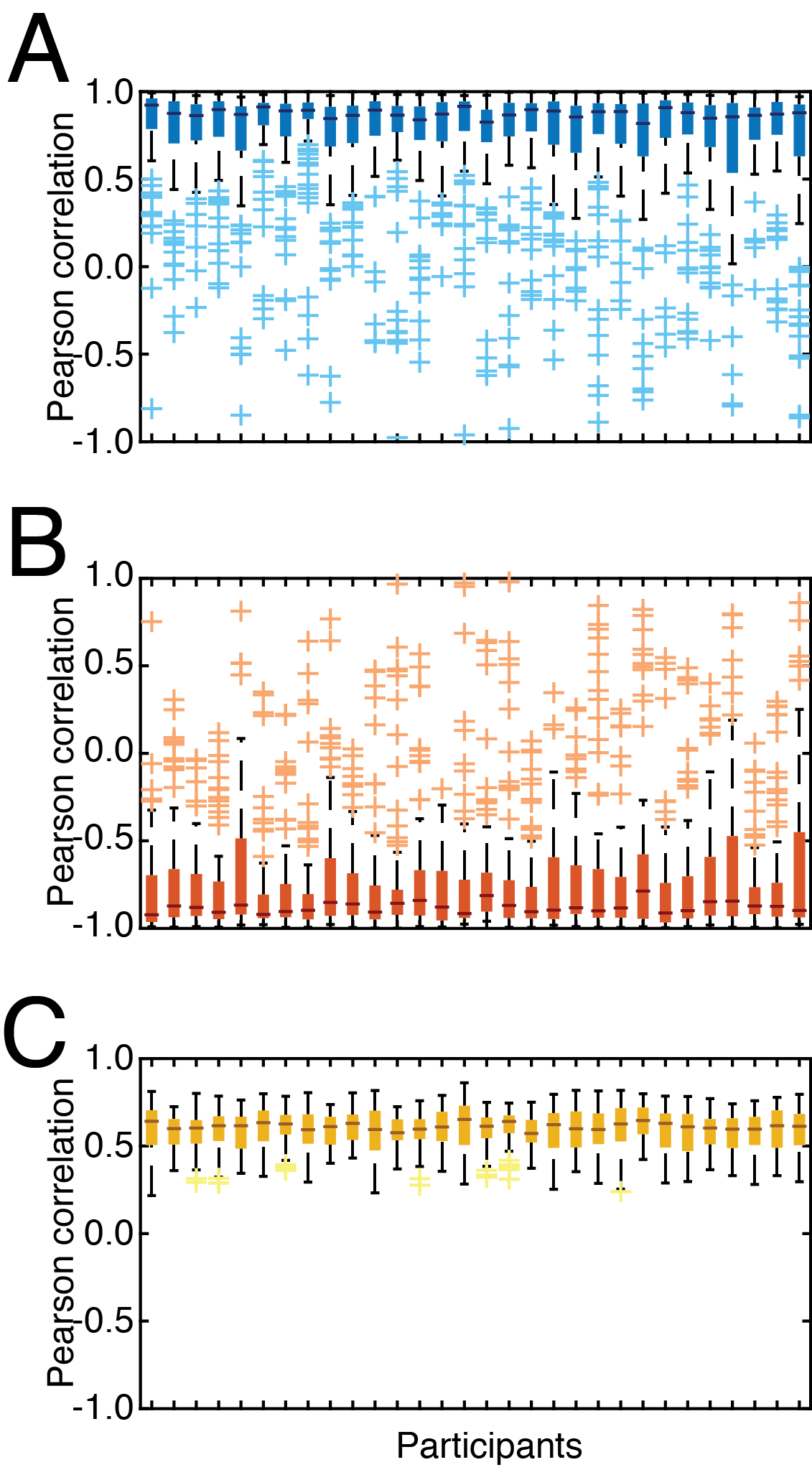}}
\caption{\textbf{Summary of full control across all participants.} In the main text we demonstrated that control energy was predicted by weighted degree (strength). Here we show that this same general pattern holds across all 30 participants. We show, here, for all three classes -- \emph{(A)} initial, \emph{(B)} target, and \emph{(C)} bulk -- the distribution of correlation coefficients (logarithm weighted degree \emph{versus} logarithm energy) obtained across all control tasks.} \label{suppfullControlAcrossSubjects}
\end{center}
\end{figure}

\begin{figure}
\begin{center}
\centerline{\includegraphics[width=0.5\textwidth]{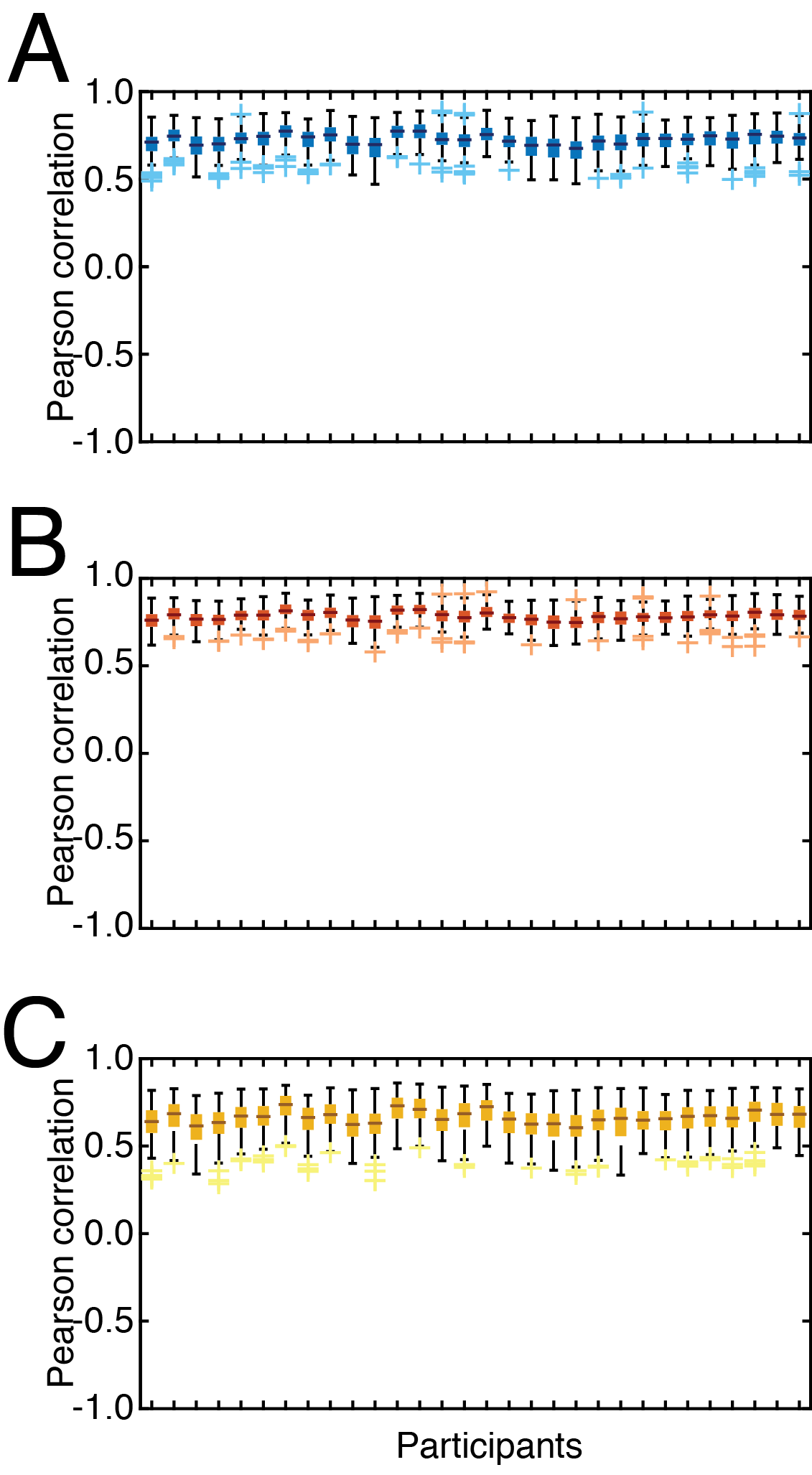}}
\caption{\textbf{Summary of single-region suppression across all participants.} In the main text we demonstrated that communicability between two regions predicted the extent to which either region compensated for the suppression of the other. We show, here, for all three classes -- \emph{(A)} initial, \emph{(B)} target, and \emph{(C)} bulk -- the distribution of correlation coefficients (communicability \emph{versus} percent change in energy) obtained across all control tasks.} \label{suppremoveSingleNodesAcrossSubjects}
\end{center}
\end{figure}

\begin{figure}
\begin{center}
\centerline{\includegraphics[width=0.5\textwidth]{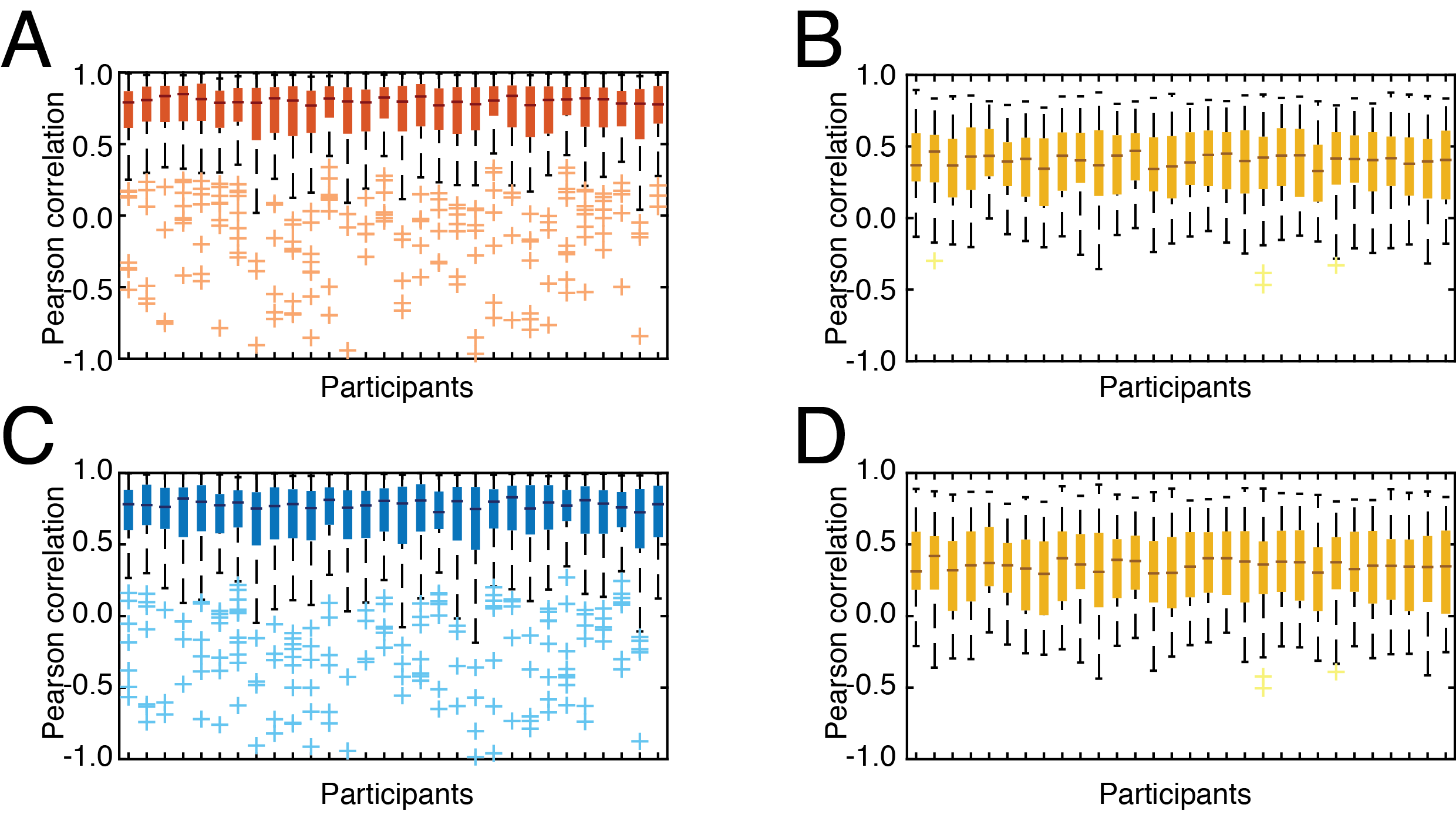}}
\caption{\textbf{Summary of initial or target class suppression across all participants.} In the main text we demonstrated that suppressing entire classes of brain regions (either initial or target classes) led to compensatory responses from the remaining brain regions. Moreover, we demonstrated that the percent change in energy of the remaining regions was closely predicted by their communicability to the suppressed class. Here, we show that this effect is consistent across all participants. Panels \emph{(A)} and \emph{(B)} show the distribution of correlation coefficients (communicability to initial class regions \emph{versus} percent change in energy) obtained for target and bulk regions when we suppressed the initial class. Panels \emph{(C)} and \emph{(D)} show the same correlation coefficients but for initial and bulk classes when the target regions were suppressed.} \label{suppremoveEachClassAcrossSubjects}
\end{center}
\end{figure}

\begin{figure}
\begin{center}
\centerline{\includegraphics[width=0.5\textwidth]{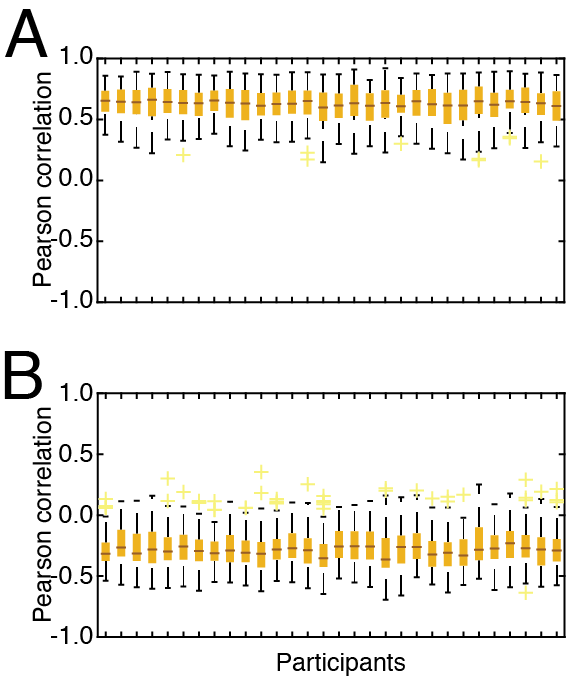}}
\caption{\textbf{Summary of initial and target class suppression across all participants.} In the main text we demonstrated that suppressing entire classes of brain regions (\emph{both} initial or target classes) led to compensatory responses from the remaining brain regions. We showed that the energy associated with the remaining regions was proportional to their communicability to both suppressed classes and that their percent change in energy was related to how far their connected neighbors were from their desired state under free evolution. Here, we recapitulate those results for each participant. Panel \emph{(A)} shows the distribution of correlation coefficients (energy \emph{versus} communicability to initial and target classes) across all control tasks for each participant. Panel \emph{(B)} shows the distribution of correlation coefficients (distance of neighbors from desired state \emph{versus} percent change in energy).} \label{suppremoveBothClassesAcrossSubjects}
\end{center}
\end{figure}

\begin{figure}
\begin{center}
\centerline{\includegraphics[width=0.5\textwidth]{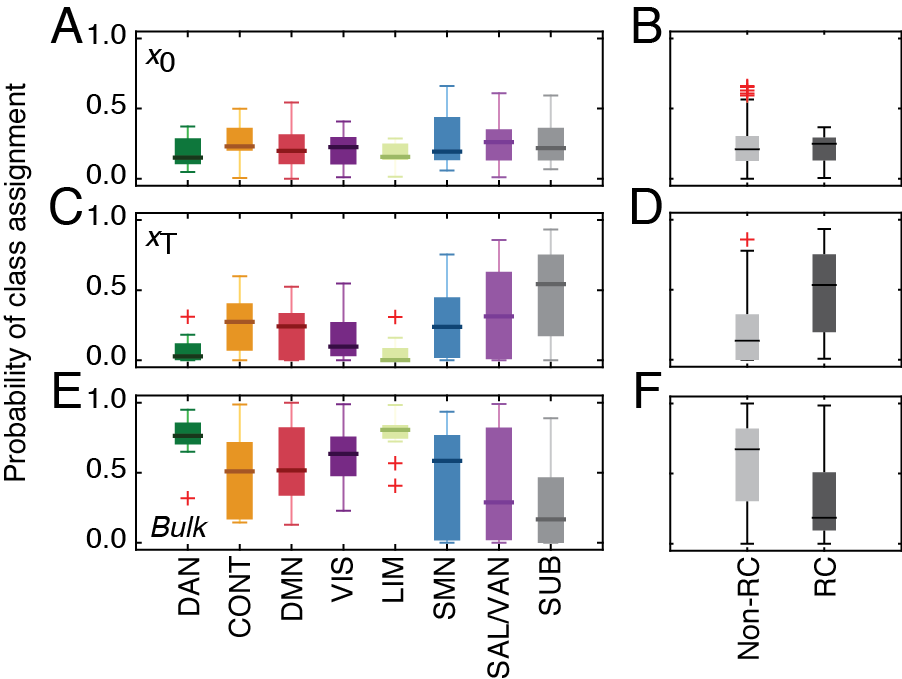}}
\caption{\textbf{Class assignments by brain system and rich club.} We aggregate class assignment probabilities according to brain system -- panels \emph{(A),(C),(E)} -- and whether or those regions were assigned to the rich club at $k=84$ -- panels \emph{(B),(D),(F)}. Each row represents a different node class: target ($\mathbf{x}_T$), initial ($\mathbf{x}_0$), and bulk classes. The probability of being assigned to the target class was statistically greater for rich club regions compared to non-rich club regions in 27/30 participants compared to 13/30 and 9/30 for bulk and initial classes, respectively ($p < 0.05$; Bonferroni corrected).} \label{suppclassAssignmentsBySystemAndRC}
\end{center}
\end{figure}

\begin{figure*}
\begin{center}
\centerline{\includegraphics[width=1\textwidth]{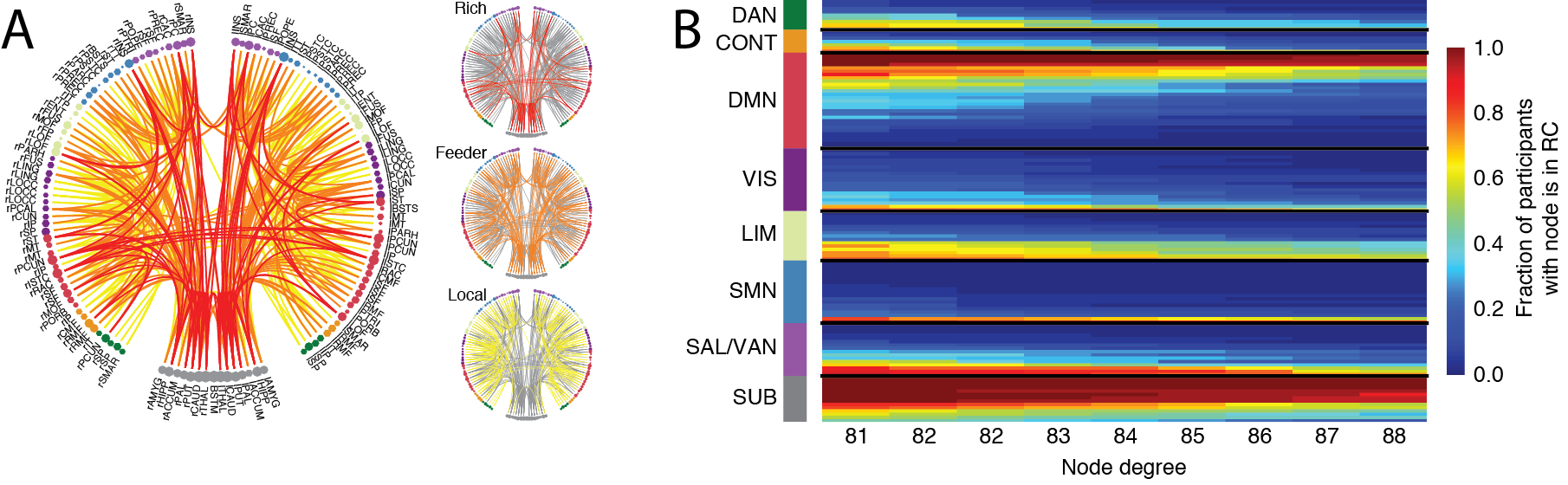}}
\caption{\textbf{Consistency of rich club.} In the main text we focus on the rich club defined at $k=84$, shown in \emph{(A)}. In \emph{(B)} we show that the composition of the rich club over the range $k=81$ to $k=88$ is, generally, consistent. Each row represents a brain region and each column a different value of $k$. The color of each cell represents the fraction of participants for which a brain region was assigned to the rich club at a given $k$.} \label{supprichClubAtDifferentK}
\end{center}
\end{figure*}

\begin{figure}
\begin{center}
\centerline{\includegraphics[width=0.5\textwidth]{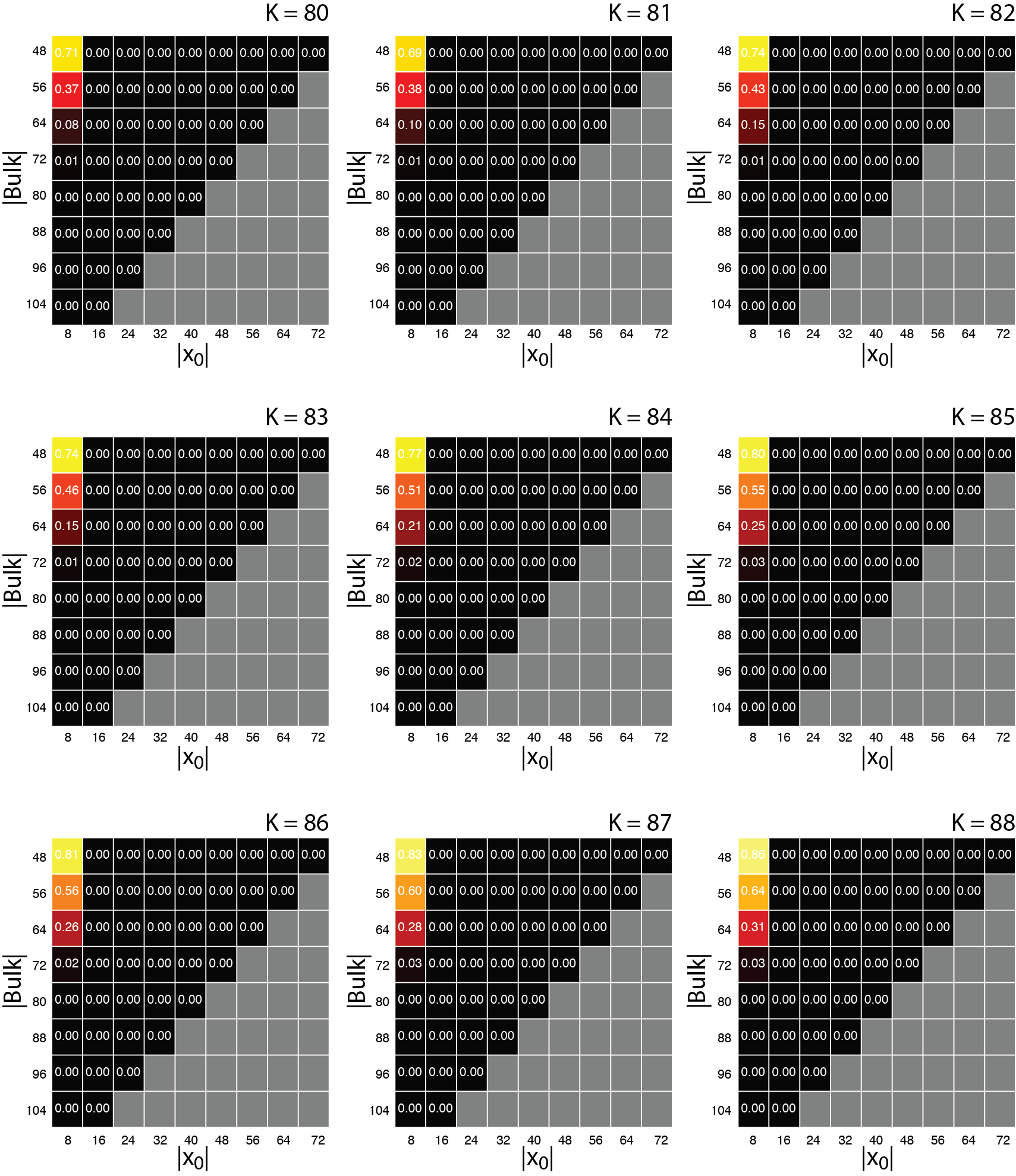}}
\caption{\textbf{Effect of rich club rewiring on energy.} In the main text we focus on the rich club defined at $k=84$, demonstrating that when rich club connections were rewired the energy associated with transitioning from the optimal initial class to the optimal target class increased. Here we show the robustness of that result with respect to variation in the level at which the rich club was defined. Each plot shows a different rich club, ranging from $k=80$ to $k=88$. The y-axis shows the number of nodes assigned to the bulk class and the x-axis shows the number of nodes assigned to the initial class. Not shown is the number of target nodes, which can be calculated as $N - |\text{bulk}| - |\mathbf{x}_0|$. Gray cells correspond to class compositions that were not possible. The remaining cells display the probability that rewiring rich club connections will reduce energy, averaged across all participants.} \label{supprichClubPvals}
\end{center}
\end{figure}

\begin{figure}
\begin{center}
\centerline{\includegraphics[width=0.5\textwidth]{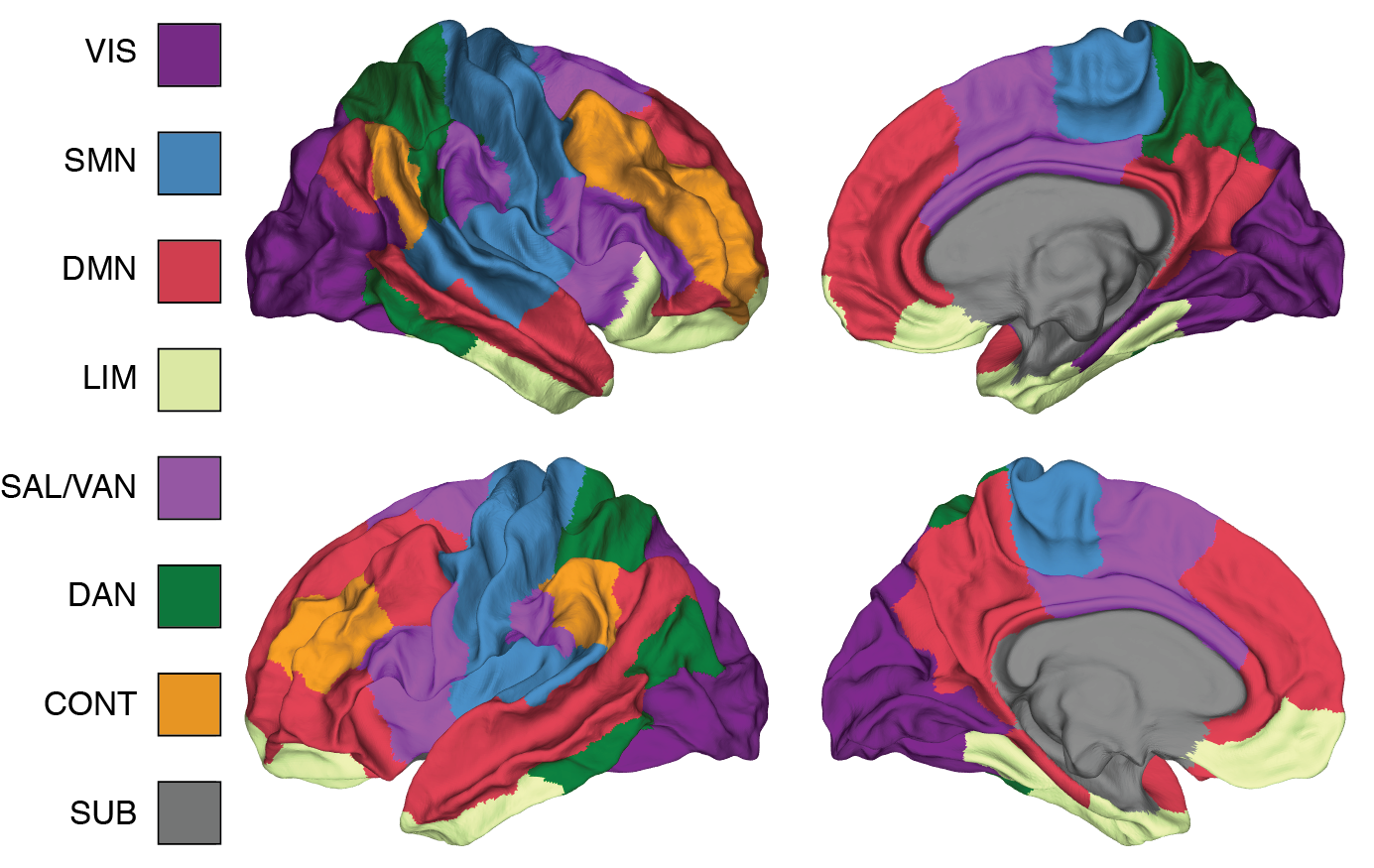}}
\caption{\textbf{Brain system assignments.} Topographic distributions of eight brain systems: Visual system (VIS; dark purple), somatomotor network (SMN; blue), default mode network (DMN; red), limbic system (LIM; cream), saliency/ventral attention network (SAL/VEN; light purple), dorsal attention network (DAN; green), control network (CONT; orange), and subcortex (SUB; slate).} \label{figure:suppBrainSystems}
\end{center}
\end{figure}

\end{document}